\documentclass[a4paper]{llncs}

\usepackage{aircraftshapes}

\usepackage{wrapfig}
\usepackage{paralist}
\usepackage{xspace}
\usepackage{amsmath}
\usepackage{amssymb}
\usepackage{url}
\usepackage{algorithm}
\usepackage{algorithmic}
 
\usepackage{hyperref}
\hypersetup{colorlinks,breaklinks,urlcolor=blue,linkcolor=blue,citecolor=blue}

\newcommand{\sat}{\texttt{SAT}}
\newcommand{\unsat}{\texttt{UNSAT}}

\newcommand{\incN}{\texttt{inc}}
\newcommand{\decN}{\texttt{dec}}
\newcommand{\posN}{\texttt{pos}}
\newcommand{\negN}{\texttt{neg}}
\newcommand{\abstractOp}{\texttt{abstract}}
\newcommand{\refineOp}{\texttt{refine}}

\newcommand{\relu}{\text{ReLU}\xspace{}}

\usepackage{tikz,ifthen,pgfplots}
\usetikzlibrary{arrows,trees,backgrounds,automata,shapes,decorations,plotmarks,fit,calc,positioning,shadows,chains}
\tikzstyle{every pin edge}=[<-,shorten <=1pt]
\tikzstyle{neuron}=[circle,fill=black!25,minimum size=17pt,inner sep=0pt]
\tikzstyle{input neuron}=[neuron, fill=orange!50]
\tikzstyle{output neuron}=[neuron, fill=purple!50]
\tikzstyle{hidden neuron}=[neuron, fill=blue!50]
\tikzstyle{annot} = [text width=3.5em, text centered]
\tikzstyle{nnedge} = [-{stealth},shorten >=0.1cm, shorten <=0.05cm,line width=0.5pt,black]

\usetikzlibrary{angles,quotes}

\begin{document}

\title{An Abstraction-Based Framework for Neural Network Verification}

\author{
  Yizhak Yisrael Elboher\inst{1} \and
  Justin Gottschlich\inst{2} \and  
  Guy Katz\inst{1} 
}
\institute{
  The Hebrew University of Jerusalem, Israel \\
  \{yizhak.elboher, g.katz\}@mail.huji.ac.il
  \and
  Intel Labs, USA \\
  justin.gottschlich@intel.com 
}

\maketitle

\begin{abstract}
  Deep neural networks are increasingly being used as controllers for
  safety-critical systems. Because neural networks are opaque,
  certifying their correctness is a significant challenge. To address
  this issue, several neural network verification approaches have
  recently been proposed. However, these approaches afford limited
  scalability, and applying them to large networks can be challenging.
  In this paper, we propose a framework that can enhance neural
  network verification techniques by using over-approximation to
  reduce the size of the network --- thus making it more amenable to
  verification. We perform the approximation such that if the property
  holds for the smaller (abstract) network, it holds for the original
  as well. The over-approximation may be too coarse, in which case the
  underlying verification tool might return a spurious
  counterexample. Under such conditions, we perform
  counterexample-guided refinement to adjust the approximation, and
  then repeat the process. Our approach is orthogonal to, and can be
  integrated with, many existing verification techniques. For
  evaluation purposes, we integrate it with the recently proposed
  Marabou framework, and observe a significant improvement in
  Marabou's performance. Our experiments demonstrate the great
  potential of our approach for verifying larger neural networks.
\end{abstract} 

\section{Introduction}

\emph{Machine programming} (MP), the automatic generation of software,
is showing early signs of fundamentally transforming the way software
is developed~\cite{gottschlich:2018:mapl}. A key ingredient employed
by MP is the \emph{deep neural network} (DNN), which has emerged as an
effective means to semi-autonomously implement many complex software
systems. DNNs are artifacts produced by \emph{machine learning}: a
user provides examples of how a system should behave, and a machine
learning algorithm generalizes these examples into a DNN capable of
correctly handling inputs that it had not seen before. Systems with
DNN components have obtained unprecedented results in fields such as
image recognition~\cite{KrSuHi12}, game
playing~\cite{SiHuMaGuSiVaScAnPaLaDi16}, natural language
processing~\cite{HiDeYuDaMoJaSeVaNgSaKi12}, computer
networks~\cite{MaNeAl17Pensieve}, and many others, often surpassing
the results obtained by similar systems that have been carefully
handcrafted. It seems evident that this trend will increase and
intensify, and that DNN components will be deployed in various
safety-critical
systems~\cite{BoDeDwFiFlGoJaMoMuZhZhZhZi16,JuLoBrOwKo16}.

DNNs are appealing in that (in some cases) they are easier to create
than handcrafted software, while still achieving excellent
results. However, their usage also raises a challenge when it comes to
certification. Undesired behavior has been observed in many
state-of-the-art DNNs. For example, in many cases slight perturbations
to correctly handled inputs can cause severe
errors~\cite{SzZaSuBrErGoFe13,KuGoBe16}.  Because many practices for
improving the reliability of hand-crafted code have yet to be
successfully applied to DNNs (e.g., code reviews, coding guidelines,
etc.), it remains unclear how to overcome the opacity of DNNs, which
may limit our ability to certify them before they are deployed.

To mitigate this, the formal methods community has begun developing
techniques for the formal verification of
DNNs
(e.g.,~\cite{GeMiDrTsChVe18,HuKwWaWu17,KaBaDiJuKo17Reluplex,WaPeWhYaJa18}). These
techniques can automatically prove that a DNN always
satisfies a prescribed property. Unfortunately, the DNN verification
problem is computationally difficult (e.g., NP-complete, even
for simple specifications and networks~\cite{KaBaDiJuKo17Reluplex}),
and becomes exponentially more difficult as network sizes
increase. Thus, despite recent advances in DNN verification
techniques, network sizes remain a severely limiting factor.

In this work, we propose a technique by which the scalability of many
existing verification techniques can be significantly increased. The
idea is to apply the well-established notion of \emph{abstraction and
refinement}~\cite{ClGrJhLuVe10CEGAR}: replace a network $N$ that is to
be verified with a much smaller, \emph{abstract} network, $\bar{N}$,
and then verify this $\bar{N}$. Because $\bar{N}$ is smaller it can be
verified more efficiently; and it is constructed in such a way that if
it satisfies the specification, the original network $N$ also
satisfies it. In the case that $\bar{N}$ does not satisfy the
specification, the verification procedure provides a counterexample
$x$. This $x$ may be a true counterexample demonstrating that the
original network $N$ violates the specification, or it may be
\emph{spurious}. If $x$ is spurious, the network $\bar{N}$ is
\emph{refined} to make it more accurate (and slightly larger), and
then the process is repeated.  A particularly useful variant of this
approach is to use the spurious $x$ to guide the refinement process,
so that the refinement step rules out $x$ as a counterexample. This
variant, known as \emph{counterexample-guided abstraction refinement}
(\emph{CEGAR})~\cite{ClGrJhLuVe10CEGAR}, has been successfully applied
in many verification contexts.

As part of our technique we propose a method for abstracting and
refining neural networks.  Our basic abstraction step \emph{merges}
two neurons into one, thus reducing the overall number of neurons by
one. This basic step can be repeated numerous times, significantly
reducing the network size. Conversely, refinement is performed by
splitting a previously merged neuron in two, increasing the network
size but making it more closely resemble the original. A key point is
that not all pairs of neurons can be merged, as this could result in a
network that is smaller but is not an over-approximation of the
original. We resolve this by first transforming the original network
into an equivalent network where each node belongs to one of four
classes, determined by its edge weights and its effect on the network's 
output; merging neurons from the same class can then be
done safely. The actual choice of which neurons to merge
or split is performed heuristically. We propose and discuss several
possible heuristics.

For evaluation purposes, we implemented our approach as a Python
framework that wraps the Marabou verification
tool~\cite{KaHuIbJuLaLiShThWuZeDiKoBa19Marabou}. We then used our
framework to verify properties of the Airborne Collision Avoidance
System (ACAS Xu) set of benchmarks~\cite{KaBaDiJuKo17Reluplex}. Our
results strongly demonstrate the potential usefulness of abstraction
in enhancing existing verification schemes: specifically, in most
cases the abstraction-enhanced Marabou significantly outperformed the
original.  Further, in most cases the properties in question
could indeed be shown to hold or not hold for the original DNN by
verifying a small, abstract version thereof.

To summarize, our contributions are:
\begin{inparaenum}[(i)]
\item
  we propose a general framework for over-approximating and refining DNNs;
\item
  we propose several heuristics for abstraction and refinement, to be used within our general framework; and
\item
  we provide an implementation of our technique that integrates
  with the Marabou verification tool and use it for
  evaluation. Our code is available online~\cite{cegarabouCode}.
\end{inparaenum}

The rest of this paper is organized as follows. In
Section~\ref{sec:background}, we provide a brief background on neural
networks and their verification. In
Section~\ref{sec:abstractionRefinement}, we describe our general
framework for abstracting an refining DNNs. In
Section~\ref{sec:cegar}, we discuss how to apply these abstraction and
refinement steps as part of a CEGAR procedure, followed by an
evaluation in Section~\ref{sec:evaluation}. In
Section~\ref{sec:relatedWork}, we discuss related work, and we conclude in
Section~\ref{sec:conclusion}.

\section{Background}
\label{sec:background}

\subsection{Neural Networks}

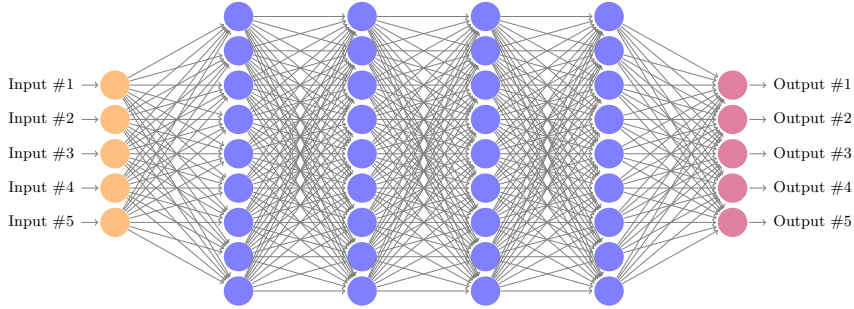
\begin{figure}
\begin{center}
\scalebox{0.65}{
\def\layersep{2.5cm}
\def\vertSepFactory{0.7}
\begin{tikzpicture}[shorten >=1pt,->,draw=black!50, node distance=\layersep]
    \foreach \name / \y in {1,...,5}
        \node[input neuron, pin=left:Input \#\y] (I-\name) at (0,-\vertSepFactory * \y) {};

    \foreach \name / \y in {1,...,9}
        \path[yshift=1.4cm]
            node[hidden neuron] (H1-\name) at (1*\layersep,-\vertSepFactory * \y cm) {};

    \foreach \name / \y in {1,...,9}
        \path[yshift=1.4cm]
            node[hidden neuron] (H2-\name) at (2*\layersep,-\vertSepFactory * \y cm) {};

    \foreach \name / \y in {1,...,9}
        \path[yshift=1.4cm]
            node[hidden neuron] (H3-\name) at (3*\layersep,-\vertSepFactory * \y cm) {};

    \foreach \name / \y in {1,...,9}
        \path[yshift=1.4cm]
            node[hidden neuron] (H4-\name) at (4*\layersep,-\vertSepFactory * \y cm) {};

    \foreach \name / \y in {1,...,5}
        \node[output neuron,pin={[pin edge={->}]right:Output \#\y}]
        (O-\name) at (5*\layersep, -\vertSepFactory * \y cm) {};

    \foreach \source in {1,...,5}
        \foreach \dest in {1,...,9}
            \path (I-\source) edge (H1-\dest);

    \foreach \source in {1,...,9}
        \foreach \dest in {1,...,9}
            \path (H1-\source) edge (H2-\dest);

    \foreach \source in {1,...,9}
        \foreach \dest in {1,...,9}
            \path (H2-\source) edge (H3-\dest);

    \foreach \source in {1,...,9}
        \foreach \dest in {1,...,9}
            \path (H3-\source) edge (H4-\dest);

    \foreach \source in {1,...,9}
        \foreach \dest in {1,...,5}
            \path (H4-\source) edge (O-\dest);

\end{tikzpicture}
}
\caption{A fully connected, feedforward DNN with 5 input nodes (in orange), 5 output
  nodes (in purple), and 4 hidden layers containing a total of 36 hidden nodes (in
  blue). Each edge is associated with a weight value (not depicted).
}
\label{fig:fullyConnectedNetwork}
\end{center}
\end{figure}

A neural network consists of an \emph{input layer}, an \emph{output
layer}, and one or more intermediate layers called \emph{hidden
layers}. Each layer is a collection of nodes, called
\emph{neurons}. Each neuron is connected to other neurons by one or
more directed edges.
In a feedforward neural network, the neurons in the first layer
receive input data that sets their initial values. The remaining
neurons calculate their values using the weighted values of the
neurons that they are connected to through edges from the preceding
layer (see Fig.~\ref{fig:fullyConnectedNetwork}). The output layer
provides the resulting value of the DNN for a given input.
 
There are many types of DNNs, which may differ in the way their neuron
values are computed. Typically, a neuron is evaluated by first
computing a weighted sum of the preceding layer's neuron values
according to the edge weights, and then applying an activation
function to this weighted sum~\cite{FoBeCu16}. We focus here on the
Rectified Linear Unit (ReLU) activation function~\cite{NaHi10}, given
as $\relu{}(x) = \max{}(0, x)$. Thus, if the weighted sum computation
yields a positive value, it is kept; and otherwise, it is replaced by
zero.

More formally, given a DNN $N$, we use $n$ to denote the number of
layers of $N$. We denote the number of nodes of layer $i$ by
$s_i$. Layers $1$ and $n$ are the input and output layers,
respectively. Layers $2,\ldots,n-1$ are the hidden layers. We denote
the value of the $j$-th node of layer $i$ by $v_{i,j}$, and denote the
column vector $[v_{i,1},\ldots,v_{i,s_i}]^T$ as $V_i$.

Evaluating $N$ is performed by calculating $V_n$ for a given input
assignment $V_1$. This is done by sequentially computing $V_i$ for
$i=2,3,\ldots,n$, each time using the values of $V_{i-1}$ to compute
weighted sums, and then applying the ReLU activation functions.
Specifically, layer $i$ (for $i>1$) is associated with a weight matrix
$W_i$ of size $s_{i}\times s_{i-1}$ and a bias vector $B_i$ of size $
s_i$. If $i$ is a hidden layer, its values are given by $ V_i = \relu{}(W_i V_{i-1} + B_i),
$ where the ReLUs are applied element-wise; and the output layer is
given by $V_n = W_nV_{n-1}+B_n$ (ReLUs are not applied). Without loss of
generality, in the rest of the paper we assume that all bias values
are 0, and can be ignored.  This rule is applied repeatedly once for
each layer, until $V_n$ is eventually computed.

We will sometimes use the notation $w(v_{i,j},v_{i+1,k})$ to refer to
the entry of $W_{i+1}$ that represents the weight of the edge between
neuron $j$ of layer $i$ and neuron $k$ of layer $i+1$. We will also
refer to such an edge as an \emph{outgoing edge} for $v_{i,j}$, and as an
\emph{incoming edge} for $v_{i+1,k}$.

As part of our abstraction framework, we will sometimes need to
consider a \emph{suffix} of a DNN, in which the first layers of the
DNN are omitted. For $1<i<n$, we use $N^{[i]}$ to denote the DNN
comprised of layers $i, i+1,\ldots,n$ of the original network. The
sizes and weights of the remaining layers are unchanged, and layer $i$
of $N$ is treated as the input layer of $N^{[i]}$.

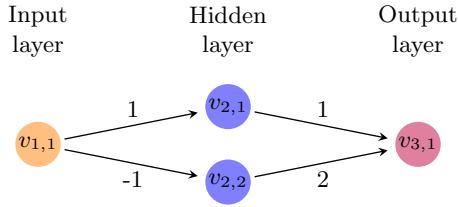
\begin{figure}[htp]
  \begin{center}
    \scalebox{1} {
      \def\layersep{2.5cm}
    \begin{tikzpicture}[shorten >=1pt,->,draw=black!50, node distance=\layersep,font=\footnotesize]

      \node[input neuron] (I-1) at (0,-1) {$v_{1,1}$};

      \path[yshift=0.5cm] node[hidden neuron] (H-1)
      at (\layersep,-1 cm) {$v_{2,1}$};
      \path[yshift=0.5cm] node[hidden neuron] (H-2)
      at (\layersep,-2 cm) {$v_{2,2}$};

      \node[output neuron] at (2*\layersep, -1) (O-1) {$v_{3,1}$};

      \draw[nnedge] (I-1) -- node[above] {$1$} (H-1);
      \draw[nnedge] (I-1) -- node[below] {-$1$} (H-2);
      \draw[nnedge] (H-1) -- node[above] {$1$} (O-1);
      \draw[nnedge] (H-2) -- node[below] {$2$} (O-1);
      
      \node[annot,above of=H-1, node distance=1cm] (hl) {Hidden layer};
      \node[annot,left of=hl] {Input layer};
      \node[annot,right of=hl] {Output layer};
    \end{tikzpicture}
    }
    \caption{A simple feedforward neural network.}
    \label{fig:runningExample}
  \end{center}
\end{figure}

Fig.~\ref{fig:runningExample} depicts a small neural network. The
network has $n=3$ layers, of sizes $s_1=1, s_2=2$ and $s_3=1$. Its
weights are $w(v_{1,1},v_{2,1})=1$, $w(v_{1,1},v_{2,2})=-1$,
$w(v_{2,1},v_{3,1})=1$ and $w(v_{2,2},v_{3,1})=2$.  For input
$v_{1,1}=3$, node $v_{2,1}$ evaluates to 3 and node $v_{2,2}$
evaluates to 0, due to the ReLU activation function. The output node
$v_{3,1}$ then evaluates to 3.

\subsection{Neural Network Verification}
DNN verification amounts to answering the following
question: given a DNN $N$, which maps input vector $x$ to output
vector $y$, and predicates $P$ and $Q$, does there exist an input
$x_0$ such that $P(x_0)$ and $Q(N(x_0))$ both hold?
In other words, the verification process determines whether there
exists a particular input that meets the input criterion $P$, and that is
mapped to an output that meets the output criterion $Q$. We refer to
$\langle N, P, Q\rangle$ as the \emph{verification query}. As is usual
in verification, $Q$ represents the \emph{negation} of the desired
property. Thus, if the query is \emph{unsatisfiable} (\unsat{}), the
property holds; and if it is \emph{satisfiable} (\sat{}), then $x_0$
constitutes a counterexample to the property in question.

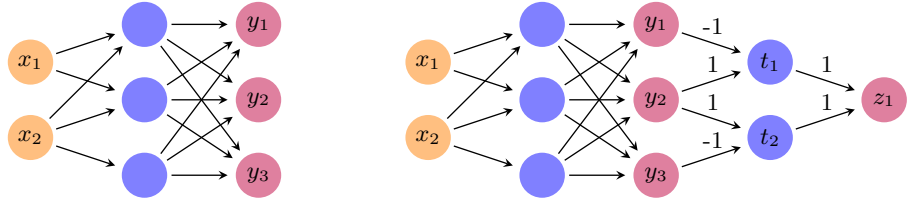
\begin{figure*}[htp]

    \begin{minipage}{.42\textwidth}

    \scalebox{1} {
      \def\layersep{1.5cm}
    \begin{tikzpicture}[shorten >=1pt,->,draw=black!50, node distance=\layersep,font=\footnotesize]

      \path[yshift=0.5cm] node[input neuron] (I-1) at (0,-1) {$x_1$};
      \path[yshift=0.5cm] node[input neuron] (I-2) at (0,-2) {$x_2$};

      \node[hidden neuron] (H-1)
      at (\layersep,0 cm) {};
      \node[hidden neuron] (H-2)
      at (\layersep,-1 cm) {};
      \node[hidden neuron] (H-3)
      at (\layersep,-2 cm) {};

      \node[output neuron] at (2*\layersep, -0) (O-1) {$y_1$};
      \node[output neuron] at (2*\layersep, -1) (O-2) {$y_2$};
      \node[output neuron] at (2*\layersep, -2) (O-3) {$y_3$};

      \draw[nnedge] (I-1) -- node[above] {$$} (H-1);
      \draw[nnedge] (I-1) -- node[above] {$$} (H-2);

      \draw[nnedge] (I-2) -- node[above] {$$} (H-1);
      \draw[nnedge] (I-2) -- node[below] {$$} (H-2);
      \draw[nnedge] (I-2) -- node[above] {$$} (H-3);

      \draw[nnedge] (H-1) -- node[above] {$$} (O-1);
      \draw[nnedge] (H-1) -- node[above] {$$} (O-2);
      \draw[nnedge] (H-1) -- node[above] {$$} (O-3);
      
      \draw[nnedge] (H-2) -- node[above] {$$} (O-1);
      \draw[nnedge] (H-2) -- node[above] {$$} (O-2);
      \draw[nnedge] (H-2) -- node[above] {$$} (O-3);

      \draw[nnedge] (H-3) -- node[above] {$$} (O-1);
      \draw[nnedge] (H-3) -- node[above] {$$} (O-2);
      \draw[nnedge] (H-3) -- node[above] {$$} (O-3);
      
    \end{tikzpicture}
  }

\end{minipage}
\begin{minipage}{.45\textwidth}

  \scalebox{1} {
      \def\layersep{1.5cm}
    \begin{tikzpicture}[shorten >=1pt,->,draw=black!50, node distance=\layersep,font=\footnotesize]

      \path[yshift=0.5cm] node[input neuron] (I-1) at (0,-1) {$x_1$};
      \path[yshift=0.5cm] node[input neuron] (I-2) at (0,-2) {$x_2$};

      \node[hidden neuron] (H-1)
      at (\layersep,0 cm) {};
      \node[hidden neuron] (H-2)
      at (\layersep,-1 cm) {};
      \node[hidden neuron] (H-3)
      at (\layersep,-2 cm) {};

      \node[output neuron] at (2*\layersep, -0) (O-1) {$y_1$};
      \node[output neuron] at (2*\layersep, -1) (O-2) {$y_2$};
      \node[output neuron] at (2*\layersep, -2) (O-3) {$y_3$};
      
      \node[hidden neuron] (T-1)
      at (3*\layersep,-0.5 cm) {$t_1$};
      \node[hidden neuron] (T-2)
      at (3*\layersep,-1.5 cm) {$t_2$};
      
      \node[output neuron] at (4*\layersep, -1) (Z-1) {$z_1$};

      \draw[nnedge] (I-1) -- node[above] {$$} (H-1);
      \draw[nnedge] (I-1) -- node[above] {$$} (H-2);

      \draw[nnedge] (I-2) -- node[above] {$$} (H-1);
      \draw[nnedge] (I-2) -- node[below] {$$} (H-2);
      \draw[nnedge] (I-2) -- node[above] {$$} (H-3);

      \draw[nnedge] (H-1) -- node[above] {$$} (O-1);
      \draw[nnedge] (H-1) -- node[above] {$$} (O-2);
      \draw[nnedge] (H-1) -- node[above] {$$} (O-3);
      
      \draw[nnedge] (H-2) -- node[above] {$$} (O-1);
      \draw[nnedge] (H-2) -- node[above] {$$} (O-2);
      \draw[nnedge] (H-2) -- node[above] {$$} (O-3);

      \draw[nnedge] (H-3) -- node[above] {$$} (O-1);
      \draw[nnedge] (H-3) -- node[above] {$$} (O-2);
      \draw[nnedge] (H-3) -- node[above] {$$} (O-3);

      \draw[nnedge] (O-1) -- node[above] {-$1$} (T-1);
      \draw[nnedge] (O-2) -- node[above] {$1$} (T-1);

      \draw[nnedge] (O-2) -- node[above] {$1$} (T-2);
      \draw[nnedge] (O-3) -- node[above] {-$1$} (T-2);

      \draw[nnedge] (T-1) -- node[above] {$1$} (Z-1);
      \draw[nnedge] (T-2) -- node[above] {$1$} (Z-1);
      
    \end{tikzpicture}
  }

\end{minipage}
    \caption{Reducing a complex property to the $y>0$ form. For the
      network on the left hand side, suppose we wish to examine the property
      $y_2>y_1 \vee y_2 > y_3$, which is a property that involves
      multiple outputs and includes a disjunction. We do this
      (right hand side network) by
      adding two neurons, $t_1$ and $t_2$, such that
      $t_1=\relu{}(y_2-y_1)$ and $t_2=\relu{}(y_2-y_3)$. Thus, $t_1>0$
    if and only if the first disjunct, $y_2>y_1$, holds; and $t_2>0$
    if and only if the second disjunct, $y_2>y_3$, holds. Finally,
    we add a neuron $z_1$ such that $z_1=t_1 + t_2$. It holds that 
    $z_1>0$ if and only if $t_1>0\vee t_2>0$. Thus, we have reduced the complex
    property into an equivalent property in the desired form.}
    \label{fig:reduceToGt}
\end{figure*}

Different verification approaches may differ in \begin{inparaenum}[(i)]
\emph{\item} the kinds of neural networks they allow (specifically, the kinds
  of activation functions in use);
\emph{\item} the kinds of input properties; and
\emph{\item} the kinds of output properties.
\end{inparaenum}
For simplicity, we focus on networks that employ the ReLU activation
function. In addition, our input properties will be conjunctions of
linear constraints on the input values. Finally, we will assume that
our networks have a single output node $y$, and that the output
property is $y > c$ for a given constant $c$. We stress that these
restrictions are for the sake of simplicity. Many properties of
interest, including those with arbitrary Boolean structure and
involving multiple neurons, can be
reduced into the above single-output setting by adding a few neurons
that encode the Boolean
structure~\cite{KaBaDiJuKo17Reluplex,RuHuKw18}; see
Fig.~\ref{fig:reduceToGt} for an example. The number of neurons to be
added is typically negligible when compared to the size of the DNN. In
particular, this is true for the ACAS Xu family of
benchmarks~\cite{KaBaDiJuKo17Reluplex}, and also for adversarial
robustness queries that use the $L_\infty$ or the $L_1$ norm as a
distance
metric~\cite{CaKaBaDi17,GoKaPaBa18,KaBaDiJuKo17}. Additionally, other
piecewise-linear activation functions, such as max-pooling layers, can
also be encoded using \relu{}s~\cite{CaKaBaDi17}.

Several techniques have been proposed for solving the aforementioned
verification problem in recent years (Section~\ref{sec:relatedWork}
includes a brief overview). Our abstraction technique is designed to
be compatible with most of these techniques, by simplifying the
network being verified, as we describe next.

\section{Network Abstraction and Refinement}
\label{sec:abstractionRefinement}
Because the complexity of verifying a neural network is strongly
connected to its size~\cite{KaBaDiJuKo17Reluplex}, our goal is to
transform a verification query $\varphi_1 = \langle N, P, Q\rangle$
into query $\varphi_2=\langle \bar{N}, P, Q\rangle$, such that the
abstract network $\bar{N}$ is significantly smaller than $N$ (notice
that properties $P$ and $Q$ remain unchanged). We will construct
$\bar{N}$ so that it is an over-approximation of $N$, meaning that if
$\varphi_2$ is \unsat{} then $\varphi_1$ is also \unsat{}. More
specifically, since our DNNs have a single output, we can regard
$N(x)$ and $\bar{N}(x)$ as real values for every input $x$. To
guarantee that $\varphi_2$ over-approximates $\varphi_1$, we will make
sure that for every $x$, $N(x)\leq \bar{N}(x)$; and thus,
$\bar{N}(x)\leq c \implies N(x)\leq c$. Because our output properties
always have the form $N(x)>c$, it is indeed the case that if
$\varphi_2$ is \unsat{}, i.e. $\bar{N}(x)\leq c$ for all $x$, then
$N(x)\leq c$ for all $x$ and so $\varphi_1$ is also \unsat{}.  We now
propose a framework for generating various $\bar{N}$s with this
property.

\subsection{Abstraction}
\label{sec:abstraction}
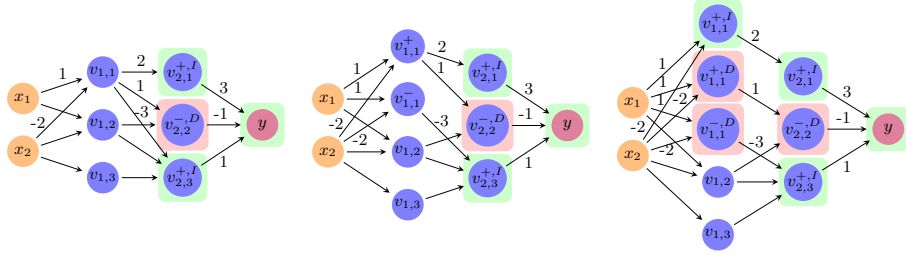
\begin{figure*}[htp]
  \begin{center}

    \begin{minipage}{.32\textwidth}

    \scalebox{0.7} {
      \def\layersep{1.5cm}
    \begin{tikzpicture}[shorten >=1pt,->,draw=black!50, node distance=\layersep,font=\footnotesize]

      \path[yshift=0.5cm] node[input neuron] (I-1) at (0,-1) {$x_1$};
      \path[yshift=0.5cm] node[input neuron] (I-2) at (0,-2) {$x_2$};

      \node[hidden neuron] (H-1)
      at (\layersep,0 cm) {$v_{1,1}$};
      \node[hidden neuron] (H-2)
      at (\layersep,-1 cm) {$v_{1,2}$};
      \node[hidden neuron] (H-3)
      at (\layersep,-2 cm) {$v_{1,3}$};

      \node[hidden neuron] (H-4)
      at (2*\layersep,-0 cm) {$v^{+,I}_{2,1}$};
      \node[hidden neuron] (H-5)
      at (2*\layersep,-1 cm) {$v^{-,D}_{2,2}$};
      \node[hidden neuron] (H-6)
      at (2*\layersep,-2 cm) {$v^{+,I}_{2,3}$};

      \node[output neuron] at (3*\layersep, -1) (O-1) {$y$};

      \draw[nnedge] (I-1) -- node[above] {$1$} (H-1);
      \draw[nnedge] (I-1) -- node[above] {$$} (H-2);

      \draw[nnedge] (I-2) -- node[left, pos=0.3] {-$2$} (H-1);
      \draw[nnedge] (I-2) -- node[below] {$$} (H-2);
      \draw[nnedge] (I-2) -- node[above] {$$} (H-3);

      \draw[nnedge] (H-1) -- node[above] {$2$} (H-4);
      \draw[nnedge] (H-1) -- node[above] {$1$} (H-5);
      \draw[nnedge] (H-1) -- node[above, pos=0.5] {-$3$} (H-6);
      \draw[nnedge] (H-2) -- node[above] {$$} (H-5);
      \draw[nnedge] (H-2) -- node[below] {$$} (H-6);
      \draw[nnedge] (H-3) -- node[below] {$$} (H-6);

      \draw[nnedge] (H-4) -- node[above] {$3$} (O-1);
      \draw[nnedge] (H-5) -- node[above, pos=0.4] {-$1$} (O-1);
      \draw[nnedge] (H-6) -- node[below] {$1$} (O-1);

      \begin{pgfonlayer}{background}
        \newcommand*{\nodePadding}{0.2cm}
        \tikzstyle{background_rectangle}=[rounded corners, fill = navy!10]

        \draw[fill = green!20, draw=none, rounded corners]
        ($(O-1.south west) + (-\nodePadding, -\nodePadding ) $)
        rectangle
        ($(O-1.north east) + ( \nodePadding, \nodePadding) $);

        \draw[fill = green!20, draw=none, rounded corners]
        ($(H-4.south west) + (-\nodePadding, -\nodePadding ) $)
        rectangle
        ($(H-4.north east) + ( \nodePadding, \nodePadding) $);

        \draw[fill = red!20, draw=none, rounded corners]
        ($(H-5.south west) + (-\nodePadding, -\nodePadding ) $)
        rectangle
        ($(H-5.north east) + ( \nodePadding, \nodePadding) $);

        \draw[fill = green!20, draw=none, rounded corners]
        ($(H-6.south west) + (-\nodePadding, -\nodePadding ) $)
        rectangle
        ($(H-6.north east) + ( \nodePadding, \nodePadding) $);

      \end{pgfonlayer}

    \end{tikzpicture}
  }

\end{minipage}
\begin{minipage}{.32\textwidth}

      \scalebox{0.7} {
      \def\layersep{1.5cm}
    \begin{tikzpicture}[shorten >=1pt,->,draw=black!50, node distance=\layersep,font=\footnotesize]

      \node[input neuron] (I-1) at (0,-1) {$x_1$};
      \node[input neuron] (I-2) at (0,-2) {$x_2$};

      \node[hidden neuron] (H-1+)
      at (\layersep,0 cm) {$v^+_{1,1}$};
      \node[hidden neuron] (H-1-)
      at (\layersep,-1 cm) {$v^-_{1,1}$};
      \node[hidden neuron] (H-2)
      at (\layersep,-2 cm) {$v_{1,2}$};
      \node[hidden neuron] (H-3)
      at (\layersep,-3 cm) {$v_{1,3}$};

      \path[yshift=0.5cm] node[hidden neuron] (H-4)
      at (2*\layersep,-1 cm) {$v^{+,I}_{2,1}$};
      \path[yshift=0.5cm] node[hidden neuron] (H-5)
      at (2*\layersep,-2 cm) {$v^{-,D}_{2,2}$};
      \path[yshift=0.5cm] node[hidden neuron] (H-6)
      at (2*\layersep,-3 cm) {$v^{+,I}_{2,3}$};

      \path[yshift=0.5cm] node[output neuron] at (3*\layersep, -2 cm) (O-1) {$y$};

      \draw[nnedge] (I-1) -- node[above, pos=0.3] {$1$} (H-1+);
      \draw[nnedge] (I-1) -- node[above, pos=0.3] {$1$} (H-1-);
      \draw[nnedge] (I-1) -- node[above] {$$} (H-2);

      \draw[nnedge] (I-2) -- node[left, pos=0.2] {-$2$} (H-1+);
      \draw[nnedge] (I-2) -- node[right, pos=0.1] {-$2$} (H-1-);
      \draw[nnedge] (I-2) -- node[below] {$$} (H-2);
      \draw[nnedge] (I-2) -- node[above] {$$} (H-3);

      \draw[nnedge] (H-1+) -- node[above, pos=0.4] {$2$} (H-4);
      \draw[nnedge] (H-1+) -- node[above, pos=0.4] {$1$} (H-5);
      \draw[nnedge] (H-1-) -- node[above, pos=0.4] {-$3$} (H-6);
      \draw[nnedge] (H-2) -- node[above] {$$} (H-5);
      \draw[nnedge] (H-2) -- node[below] {$$} (H-6);
      \draw[nnedge] (H-3) -- node[below] {$$} (H-6);

      \draw[nnedge] (H-4) -- node[above] {$3$} (O-1);
      \draw[nnedge] (H-5) -- node[above, pos=0.4] {-$1$} (O-1);
      \draw[nnedge] (H-6) -- node[below] {$1$} (O-1);

      \begin{pgfonlayer}{background}
        \newcommand*{\nodePadding}{0.2cm}
        \tikzstyle{background_rectangle}=[rounded corners, fill = navy!10]

        \draw[fill = green!20, draw=none, rounded corners]
        ($(O-1.south west) + (-\nodePadding, -\nodePadding ) $)
        rectangle
        ($(O-1.north east) + ( \nodePadding, \nodePadding) $);

        \draw[fill = green!20, draw=none, rounded corners]
        ($(H-4.south west) + (-\nodePadding, -\nodePadding ) $)
        rectangle
        ($(H-4.north east) + ( \nodePadding, \nodePadding) $);

        \draw[fill = red!20, draw=none, rounded corners]
        ($(H-5.south west) + (-\nodePadding, -\nodePadding ) $)
        rectangle
        ($(H-5.north east) + ( \nodePadding, \nodePadding) $);

        \draw[fill = green!20, draw=none, rounded corners]
        ($(H-6.south west) + (-\nodePadding, -\nodePadding ) $)
        rectangle
        ($(H-6.north east) + ( \nodePadding, \nodePadding) $);

      \end{pgfonlayer}

    \end{tikzpicture}
  }

\end{minipage}
\begin{minipage}{.32\textwidth}

      \scalebox{0.7} {
      \def\layersep{1.6cm}
    \begin{tikzpicture}[shorten >=1pt,->,draw=black!50, node distance=\layersep,font=\footnotesize]

      \node[input neuron] (I-1) at (0,-1) {$x_1$};
      \node[input neuron] (I-2) at (0,-2) {$x_2$};

      \path[yshift=0.5cm] node [hidden neuron] (H-1+inc)
      at (\layersep,0 cm) {$v^{+,I}_{1,1}$};
      \path[yshift=0.5cm] node [hidden neuron] (H-1+dec)
      at (\layersep,-1 cm) {$v^{+,D}_{1,1}$};
      \path[yshift=0.5cm] node[hidden neuron] (H-1-)
      at (\layersep,-2 cm) {$v^{-,D}_{1,1}$};
      \path[yshift=0.5cm] node[hidden neuron] (H-2)
      at (\layersep,-3 cm) {$v_{1,2}$};
      \path[yshift=0.5cm] node[hidden neuron] (H-3)
      at (\layersep,-4 cm) {$v_{1,3}$};

      \path[yshift=0.5cm] node[hidden neuron] (H-4)
      at (2*\layersep,-1 cm) {$v^{+,I}_{2,1}$};
      \path[yshift=0.5cm] node[hidden neuron] (H-5)
      at (2*\layersep,-2 cm) {$v^{-,D}_{2,2}$};
      \path[yshift=0.5cm] node[hidden neuron] (H-6)
      at (2*\layersep,-3 cm) {$v^{+,I}_{2,3}$};

      \path[yshift=0.5cm] node[output neuron] at (3*\layersep, -2 cm) (O-1) {$y$};

      \draw[nnedge] (I-1) -- node[above, pos=0.3] {$1$} (H-1+inc);
      \draw[nnedge] (I-1) -- node[above, pos=0.3] {$1$} (H-1+dec);
      \draw[nnedge] (I-1) -- node[above, pos=0.25] {$1$} (H-1-);
      \draw[nnedge] (I-1) -- node[above] {$$} (H-2);

      \draw[nnedge] (I-2) -- node[left, pos=0.15] {-$2$} (H-1+inc);
      \draw[nnedge] (I-2) -- node[left, pos=0.85] {-$2$} (H-1+dec);
      \draw[nnedge] (I-2) -- node[right, pos=0.1] {-$2$} (H-1-);
      \draw[nnedge] (I-2) -- node[below] {$$} (H-2);
      \draw[nnedge] (I-2) -- node[above] {$$} (H-3);

      \draw[nnedge] (H-1+inc) -- node[above, pos=0.4] {$2$} (H-4);
      \draw[nnedge] (H-1+dec) -- node[above, pos=0.4] {$1$} (H-5);
      \draw[nnedge] (H-1-) -- node[above, pos=0.4] {-$3$} (H-6);
      \draw[nnedge] (H-2) -- node[above] {$$} (H-5);
      \draw[nnedge] (H-2) -- node[below] {$$} (H-6);
      \draw[nnedge] (H-3) -- node[below] {$$} (H-6);

      \draw[nnedge] (H-4) -- node[above] {$3$} (O-1);
      \draw[nnedge] (H-5) -- node[above, pos=0.4] {-$1$} (O-1);
      \draw[nnedge] (H-6) -- node[below] {$1$} (O-1);

      \begin{pgfonlayer}{background}
        \newcommand*{\nodePadding}{0.2cm}
        \tikzstyle{background_rectangle}=[rounded corners, fill = navy!10]

        \draw[fill = green!20, draw=none, rounded corners]
        ($(O-1.south west) + (-\nodePadding, -\nodePadding ) $)
        rectangle
        ($(O-1.north east) + ( \nodePadding, \nodePadding) $);

        \draw[fill = green!20, draw=none, rounded corners]
        ($(H-4.south west) + (-\nodePadding, -\nodePadding ) $)
        rectangle
        ($(H-4.north east) + ( \nodePadding, \nodePadding) $);

        \draw[fill = red!20, draw=none, rounded corners]
        ($(H-5.south west) + (-\nodePadding, -\nodePadding ) $)
        rectangle
        ($(H-5.north east) + ( \nodePadding, \nodePadding) $);

        \draw[fill = green!20, draw=none, rounded corners]
        ($(H-6.south west) + (-\nodePadding, -\nodePadding ) $)
        rectangle
        ($(H-6.north east) + ( \nodePadding, \nodePadding) $);

        \draw[fill = green!20, draw=none, rounded corners]
        ($(H-1+inc.south west) + (-\nodePadding, -\nodePadding ) $)
        rectangle
        ($(H-1+inc.north east) + ( \nodePadding, \nodePadding) $);

        \draw[fill = red!20, draw=none, rounded corners]
        ($(H-1+dec.south west) + (-\nodePadding, -\nodePadding ) $)
        rectangle
        ($(H-1+dec.north east) + ( \nodePadding, \nodePadding) $);

        \draw[fill = red!20, draw=none, rounded corners]
        ($(H-1-.south west) + (-\nodePadding, -\nodePadding ) $)
        rectangle
        ($(H-1-.north east) + ( \nodePadding, \nodePadding) $);

      \end{pgfonlayer}

    \end{tikzpicture}
  }
\end{minipage}
    \caption{Classifying neurons as \posN{}/\negN{} and
      \incN{}/\decN{}. In the initial network (left), the
      neurons of the second hidden layer are already classified: $^+$
      and $^-$
    superscripts indicate \posN{} and \negN{} neurons, respectively;
    the $^I$ superscript and green background indicate \incN{}, and
    the $^D$ superscript and red background indicate \decN{}.
Classifying node $v_{1,1}$ is done by first
  splitting it into two nodes $v^+_{1,1}$ and $v^-_{1,1}$ (middle). Both nodes have identical incoming edges, but the
  outgoing edges of $v_{1,1}$ are partitioned between them, according
  to the sign of each edge's weight.
  In the last network (right), $v^+_{1,1}$ is split once
  more, into an \incN{} node with outgoing edges only to other \incN{}
  nodes, and a \decN{} node with outgoing edges only to other \decN{}
  nodes. Node $v_{1,1}$ is thus transformed into three nodes, each of which can
  finally  be classified as \incN{} or \decN{}. Notice that in the worst case, each node is split into four nodes, although for $v_{1,1}$ three nodes were enough.}
    \label{fig:positiveAndNegative}
  \end{center}
\end{figure*}

We seek to define an abstraction operator that removes a single neuron
from the network, by merging it with another neuron. To do this, we
will first transform $N$ into an equivalent network, whose neurons
have properties that will facilitate their merging. Equivalent here
means that for every input vector, both networks produce the exact
same output.  First, each hidden neuron $v_{i,j}$ of our transformed
network will be classified as either a \posN{} neuron or a \negN{}
neuron. A neuron is \posN{} if all the weights on its outgoing edges
are positive, and is \negN{} if all those weights are negative.
Second, orthogonally to the \posN{}/\negN{} classification, each
hidden neuron will also be classified as either an \incN{} neuron or a
\decN{} neuron.  $v_{i,j}$ is an \incN{} neuron of $N$ if, when we
look at $N^{[i]}$ (where $v_{i,j}$ is an input neuron), increasing the
value of $v_{i,j}$ increases the value of the network's
output. Formally, $v_{i,j}$ is \incN{} if for every two input vectors
$x_1$ and $x_2$ where $x_1[k]=x_2[k]$ for $k\neq j$ and
$x_1[j]>x_2[j]$, it holds that $N^{[i]}(x_1) > N^{[i]}(x_2)$. A
\decN{} neuron is defined symmetrically, so that \emph{decreasing} the
value of $x[j]$ \emph{increases} the output. We first describe this
transformation (an illustration of which appears in
Fig.~\ref{fig:positiveAndNegative}), and later we explain how it fits
into our abstraction framework.

Our first step is to transform $N$ into a new network, $N'$, in which
every hidden neuron is classified as \posN{} or \negN{}.  This
transformation is done by replacing each hidden neuron $v_{i_j}$ with
two neurons, $v^+_{i,j}$ and $v^-_{i,j}$, which are respectively
\posN{} and \negN{}.  Both $v^+_{i,j}$ an $v^-_{i,j}$ retain a copy of
all incoming edges of the original $v_{i,j}$; however, $v^+_{i,j}$
retains just the outgoing edges with positive weights, and $v^-_{i,j}$
retains just those with negative weights. Outgoing edges with negative
weights are removed from $v^+_{i,j}$ by setting their weights to $0$,
and the same is done for outgoing edges with positive weights for
$v^-_{i,j}$.  Formally, for every neuron $v_{i-1,p}$,
\[
  w'(v_{i-1,p},v^+_{i,j}) = w(v_{i-1,p},v_{i,j}),
  \qquad
w'(v_{i-1,p},v^-_{i,j}) = w(v_{i-1,p},v_{i,j})
\]
where $w'$ represents the weights in the new network $N'$.
Also,  for every neuron $v_{i+1,q}$
\[
w'(v^+_{i,j},v_{i+1,q}) =
\begin{cases}
      w(v_{i,j},v_{i+1,q}) & w(v_{i,j},v_{i+1,q})\geq 0 \\
      0 & \text{otherwise}
\end{cases}
\]
and
\[
w'(v^-_{i,j},v_{i+1,q}) =
\begin{cases}
      w(v_{i,j},v_{i+1,q}) & w(v_{i,j},v_{i+1,q})\leq 0 \\
      0 & \text{otherwise}
\end{cases}
\]
(see Fig.~\ref{fig:positiveAndNegative}). This operation is performed
once for every hidden neuron of $N$, resulting in a network $N'$ that
is roughly double the size of $N$. Observe that $N'$ is indeed equivalent to $N$, i.e. their outputs are always identical.

Our second step is to alter $N'$ further, into a new network $N''$,
where every hidden neuron is either \incN{} or \decN{}  (in
addition to already being \posN{} or \negN{}).
Generating $N''$ from $N'$ is performed by traversing the layers of 
$N'$ backwards, each time handling a single layer and
possibly doubling its number of neurons:
\begin{itemize}
\item Initial step: the output layer has a single neuron, $y$. This
  neuron is an \incN{} node, because increasing its value
  will increase the network's output value.
\item Iterative step: observe layer $i$, and suppose the nodes of
  layer $i+1$ have already been partitioned into \incN{} and \decN{}
  nodes. Observe a neuron $v^+_{i,j}$ in layer $i$ which is marked \posN{}
  (the case for \negN{} is symmetrical). We replace
  $v^+_{i,j}$ with two neurons $v^{+,I}_{i,j}$ and $v^{+,D}_{i,j}$,
  which are \incN{} and \decN{}, respectively.  Both new neurons
  retain a copy of all incoming edges of $v^{+}_{i,j}$; however,
  $v^{+,I}_{i,j}$ retains only outgoing edges that lead to \incN{}
  nodes, and $v^{+,D}_{i,j}$ retains only outgoing edges that lead to
  \decN{} nodes. Thus, for every $v_{i-1,p}$ and $v_{i+1,q}$,
  \[
    w''(v_{i-1,p},v^{+,I}_{i,j}) = w'(v_{i-1,p},v^+_{i,j}),
    \qquad
  w''(v_{i-1,p},v^{+,D}_{i,j}) = w'(v_{i-1,p},v^+_{i,j})
  \]
\[
w''(v^{+,I}_{i,j},v_{i+1,q}) =
\begin{cases}
  w'(v^+_{i,j},v_{i+1,q}) & \text{if } v_{i+1,q} \text{ is \incN{}} \\
  0 & \text{otherwise}
\end{cases}
\]
\[
w''(v^{+,D}_{i,j},v_{i+1,q}) =
\begin{cases}
  w'(v^+_{i,j},v_{i+1,q}) & \text{if } v_{i+1,q} \text{ is \decN{}} \\
  0 & \text{otherwise}
\end{cases}
\]
where $w''$ represents the weights in the new network $N''$.
We perform this step for each neuron in layer $i$,
resulting in neurons that are each classified as either \incN{} or
\decN{}.
\end{itemize}
To understand the intuition behind this classification, recall that by
our assumption all hidden nodes use the ReLU
activation function, which is monotonically increasing. Because $v^{+}_{i,j}$ is
\posN{}, all its outgoing edges have positive weights, and so if its
assignment was to increase (decrease), the assignments of all nodes to
which it is connected in the following layer would also increase
(decrease). Thus, we split $v^{+}_{i,j}$ in two, and make sure one
copy, $v^{+,I}_{i,j}$, is only connected to nodes that need to
increase (\incN{} nodes), and that the other copy, $v^{+,D}_{i,j}$, is
only connected to nodes that need to decrease (\decN{} nodes). This
ensures that $v^{+,I}_{i,j}$ is itself \incN{}, and that
$v^{+,D}_{i,j}$ is \decN{}. Also, both $v^{+,I}_{i,j}$ and
$v^{+,D}_{i,j}$ remain \posN{} nodes, because their outgoing edges all
have positive weights.

When this procedure terminates,  $N''$ is equivalent to
$N'$, and so also to $N$; and $N''$ is roughly double the size of $N'$, and 
roughly four times the size of $N$.  Both transformation steps
are only performed for hidden neurons, whereas the input and output
neurons remain unchanged.  This is summarized by the
following lemma:
\begin{lemma}
  \label{lemma:incDec}
Any DNN $N$ can be transformed into an equivalent network $N''$ where
each hidden neuron is \posN{} or \negN{}, and also \incN{} or \decN{},
by increasing its number of neurons by a factor of at most $4$.
\end{lemma}

Using Lemma~\ref{lemma:incDec}, we can assume without loss of
generality that the DNN nodes in our input query $\varphi_1$ are each
marked as \posN{}/\negN{} and as \incN{}/\decN{}. We are now ready to
construct the over-approximation network $\bar{N}$. We do this by
specifying an \abstractOp{} operator that merges a pair of neurons in
the network (thus reducing network size by one), and can be applied
multiple times. The only restrictions are that the two neurons being
merged need to be from the same hidden layer, and must share the same
\posN{}/\negN{} and \incN{}/\decN{} attributes. Consequently, applying
\abstractOp{} to saturation will result in a network with at most 4
neurons in each hidden layer, which over-approximates the original
network.  This, of course, would be an immense reduction in the number
of neurons for most reasonable input networks.

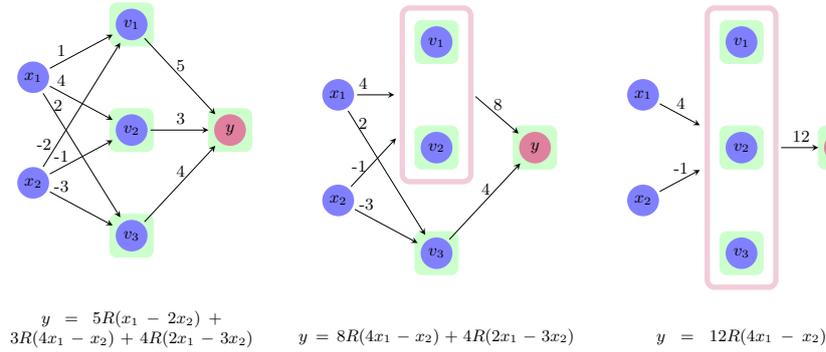
\begin{figure*}[t]
  \begin{center}
    \hspace{-0.5cm}
    \begin{minipage}{.32\textwidth}
  \center
    \scalebox{0.7} {
      \def\layersep{1.85cm}
    \begin{tikzpicture}[shorten >=1pt,->,draw=black!50, node distance=\layersep,font=\footnotesize]


      \path[] node[hidden neuron] (I-1) at (0,-1) {$x_1$};
      \path[] node[hidden neuron] (I-2) at (0,-3) {$x_2$};

      \node[hidden neuron] (H-1)
      at (\layersep, 0 cm) {$v_{1}$};
      \node[hidden neuron] (H-2)
      at (\layersep,-2 cm) {$v_{2}$};
      \node[hidden neuron] (H-3)
      at (\layersep,-4 cm) {$v_{3}$};

      \node[output neuron] at (2*\layersep, -2) (O-1) {$y$};
 
      \draw[nnedge] (I-1) -- node[above, pos=0.2] {$1$} (H-1);
      \draw[nnedge] (I-1) -- node[above, pos=0.2] {$4$} (H-2);
      \draw[nnedge] (I-1) -- node[above, pos=0.2] {$2$} (H-3);

      \draw[nnedge] (I-2) -- node[above, pos=0.1] {-$2\ \ $} (H-1);
      \draw[nnedge] (I-2) -- node[above, pos=0.2] {-$1$} (H-2);
      \draw[nnedge] (I-2) -- node[above, pos=0.2] {-$3$} (H-3);

      \draw[nnedge] (H-1) -- node[above] {$5$} (O-1);
      \draw[nnedge] (H-2) -- node[above] {$3$} (O-1);
      \draw[nnedge] (H-3) -- node[above] {$4$} (O-1);

      \node[annot,text width=6cm,below = 1cm of H-3]
           {$y=5R(x_1-2x_2) + 3R(4x_1-x_2) +4R(2x_1-3x_2)$};

      \begin{pgfonlayer}{background}
        \newcommand*{\nodePadding}{0.2cm}
        \tikzstyle{background_rectangle}=[rounded corners, fill = navy!10]

        \draw[fill = green!20, draw=none, rounded corners]
        ($(H-1.south west) + (-\nodePadding, -\nodePadding ) $)
        rectangle
        ($(H-1.north east) + ( \nodePadding, \nodePadding) $);

        \draw[fill = green!20, draw=none, rounded corners]
        ($(H-2.south west) + (-\nodePadding, -\nodePadding ) $)
        rectangle
        ($(H-2.north east) + ( \nodePadding, \nodePadding) $);

        \draw[fill = green!20, draw=none, rounded corners]
        ($(H-3.south west) + (-\nodePadding, -\nodePadding ) $)
        rectangle
        ($(H-3.north east) + ( \nodePadding, \nodePadding) $);

        \draw[fill = green!20, draw=none, rounded corners]
        ($(O-1.south west) + (-\nodePadding, -\nodePadding ) $)
        rectangle
        ($(O-1.north east) + ( \nodePadding, \nodePadding) $);

      \end{pgfonlayer}

    \end{tikzpicture}
  }

\end{minipage}
\begin{minipage}{.32\textwidth}
  \center
      \scalebox{0.7} {
      \def\layersep{1.85cm}
    \begin{tikzpicture}[shorten >=1pt,->,draw=black!50, node distance=\layersep,font=\footnotesize]

      \path[] node[hidden neuron] (I-1) at (0,-1) {$x_1$};
      \path[] node[hidden neuron] (I-2) at (0,-3) {$x_2$};

      \node[hidden neuron] (H-1)
      at (\layersep, 0 cm) {$v_{1}$};
      \node[hidden neuron] (H-2)
      at (\layersep,-2 cm) {$v_{2}$};
      \node[hidden neuron] (H-3)
      at (\layersep,-4 cm) {$v_{3}$};

      \node[fit={($(H-2.south west) + (-0.32cm, -0.32cm ) $) ($(H-1.north east) + ( 0.32cm, 0.32cm) $)}, fill = none, draw=purple!20, rounded corners, line width = 0.1cm] (absNeuron) {};

      \node[output neuron] at (2*\layersep, -2) (O-1) {$y$};

      \draw[nnedge] (I-1) -- node[above, pos=0.2] {$2$} (H-3);

      \draw[nnedge] (I-2) -- node[above, pos=0.2] {-$3$} (H-3);

      \draw[nnedge] (H-3) -- node[above] {$4$} (O-1);

      \draw[nnedge] (I-1) -- node[above, pos=0.2] {$4$} (absNeuron);
      \draw[nnedge] (I-2) -- node[above, pos=0.2] {-$1$} (absNeuron);
      \draw[nnedge] (absNeuron.east) -- node[above] {$8$} (O-1);

      \node[annot,text width=6cm,below = 1cm of H-3]
           {$y=8R(4x_1-x_2) + 4R(2x_1-3x_2)$};

      \begin{pgfonlayer}{background}
        \newcommand*{\nodePadding}{0.2cm}
        \tikzstyle{background_rectangle}=[rounded corners, fill = navy!10]

        \draw[fill = green!20, draw=none, rounded corners]
        ($(H-1.south west) + (-\nodePadding, -\nodePadding ) $)
        rectangle
        ($(H-1.north east) + ( \nodePadding, \nodePadding) $);

        \draw[fill = green!20, draw=none, rounded corners]
        ($(H-2.south west) + (-\nodePadding, -\nodePadding ) $)
        rectangle
        ($(H-2.north east) + ( \nodePadding, \nodePadding) $);

        \draw[fill = green!20, draw=none, rounded corners]
        ($(H-3.south west) + (-\nodePadding, -\nodePadding ) $)
        rectangle
        ($(H-3.north east) + ( \nodePadding, \nodePadding) $);

        \draw[fill = green!20, draw=none, rounded corners]
        ($(O-1.south west) + (-\nodePadding, -\nodePadding ) $)
        rectangle
        ($(O-1.north east) + ( \nodePadding, \nodePadding) $);

      \end{pgfonlayer}

    \end{tikzpicture}
  }

\end{minipage}
\begin{minipage}{.32\textwidth}
  \center
  \scalebox{0.7} {
      \def\layersep{1.85cm}
    \begin{tikzpicture}[shorten >=1pt,->,draw=black!50, node distance=\layersep,font=\footnotesize]

      \path[] node[hidden neuron] (I-1) at (0,-1) {$x_1$};
      \path[] node[hidden neuron] (I-2) at (0,-3) {$x_2$};

      \node[hidden neuron] (H-1)
      at (\layersep, 0 cm) {$v_{1}$};
      \node[hidden neuron] (H-2)
      at (\layersep,-2 cm) {$v_{2}$};
      \node[hidden neuron] (H-3)
      at (\layersep,-4 cm) {$v_{3}$};

      \node[fit={($(H-3.south west) + (-0.32cm, -0.32cm ) $) ($(H-1.north east) + ( 0.32cm, 0.32cm) $)}, fill = none, draw=purple!20, rounded corners, line width = 0.1cm] (absNeuron) {};

      \node[output neuron] at (2*\layersep, -2) (O-1) {$y$};


      \draw[nnedge] (I-1) -- node[above] {$4$} (absNeuron);
      \draw[nnedge] (I-2) -- node[above] {-$1$} (absNeuron);
      \draw[nnedge] (absNeuron.east) -- node[above] {$12$} (O-1);

      \node[annot,text width=6cm,below = 1cm of H-3]
           {$y=12R(4x_1-x_2)$};

      \begin{pgfonlayer}{background}
        \newcommand*{\nodePadding}{0.2cm}
        \tikzstyle{background_rectangle}=[rounded corners, fill = navy!10]

        \draw[fill = green!20, draw=none, rounded corners]
        ($(H-1.south west) + (-\nodePadding, -\nodePadding ) $)
        rectangle
        ($(H-1.north east) + ( \nodePadding, \nodePadding) $);

        \draw[fill = green!20, draw=none, rounded corners]
        ($(H-2.south west) + (-\nodePadding, -\nodePadding ) $)
        rectangle
        ($(H-2.north east) + ( \nodePadding, \nodePadding) $);

        \draw[fill = green!20, draw=none, rounded corners]
        ($(H-3.south west) + (-\nodePadding, -\nodePadding ) $)
        rectangle
        ($(H-3.north east) + ( \nodePadding, \nodePadding) $);

        \draw[fill = green!20, draw=none, rounded corners]
        ($(O-1.south west) + (-\nodePadding, -\nodePadding ) $)
        rectangle
        ($(O-1.north east) + ( \nodePadding, \nodePadding) $);

      \end{pgfonlayer}

    \end{tikzpicture}
  }
\end{minipage}
\caption{ Using \abstractOp{} to merge
  $\langle$\posN{},\incN{}$\rangle$ nodes. Initially (left), the
  three nodes $v_1,v_2$ and $v_3$ are separate. Next (middle),
  \abstractOp{} merges $v_1$ and $v_2$ into a single node. For the
  edge between $x_1$ and the new abstract node we pick the weight $4$,
  which is the maximal weight among edges from $x_1$ to $v_1$ and
  $v_2$. Likewise, the edge between $x_2$ and the abstract node has
  weight $-1$. The outgoing edge from the abstract node to $y$ has
  weight $8$, which is the sum of the weights of edges from $v_1$ and
  $v_2$ to $y$. Next, \abstractOp{} is applied again to merge $v_3$
  with the abstract node, and the weights are adjusted
  accordingly (right). With every abstraction, the value of $y$ (given as a
  formula at the bottom of each DNN, where $R$ represents the \relu{}
  operator) increases. For example, to see that
  $12R(4x_1-x_2)\geq 8R(4x_1-x_2) + 4R(2x_1-3x_2)$, it is enough to
  see that $4R(4x_1-x_2)\geq 4R(2x_1-3x_2)$, which holds because
  \relu{} is a monotonically increasing function and $x_1$ and $x_2$
  are non-negative (being, themselves, the output of \relu{} nodes).
}
\label{fig:abstraction}
  \end{center}
\end{figure*}

The \abstractOp{} operator's behavior depends on the attributes of the
neurons being merged. For simplicity, we will focus on the
$\langle$\posN{},\incN{}$\rangle$ case.  Let $v_{i,j}$, $v_{i,k}$ be
two hidden neurons of layer $i$, both classified as
$\langle$\posN{},\incN{}$\rangle$. Because layer $i$ is hidden, we
know that layers $i+1$ and $i-1$ are defined. Let $v_{i-1,p}$ and
$v_{i+1,q}$ denote arbitrary neurons in the preceding and succeeding
layer, respectively.  We construct a network $\bar{N}$ that is
identical to $N$, except that:
\begin{inparaenum}[(i)]
  \item nodes $v_{i,j}$ and $v_{i,k}$ are removed and replaced with a
    new single node, $v_{i,t}$; and
  \item all edges that touched nodes $v_{i,j}$ or $v_{i,k}$ are
    removed, and other edges are untouched.
\end{inparaenum}
Finally, we add new incoming and outgoing edges for the new node $v_{i,t}$ as
follows:
\begin{itemize}
\item Incoming edges:
$
  \bar{w}(v_{i-1,p},v_{i,t}) = \max
  \{w(v_{i-1,p},v_{i,j}), w(v_{i-1,p},v_{i,k})\}
$

\item Outgoing edges:
$
  \bar{w}(v_{i,t},v_{i+1,q}) =
  w(v_{i,j},v_{i+1,q}) + w(v_{i,k},v_{i+1,q})
$
\end{itemize}
where $\bar{w}$ represents the weights in the new network $\bar{N}$.
An illustrative example appears in
Fig.~\ref{fig:abstraction}. Intuitively, this definition of
\abstractOp{} seeks to ensure that the new node $v_{i,t}$ always
contributes more to the network's output than the two original nodes
$v_{i,j}$ and $v_{i,k}$ --- so that the new network produces a larger
output than the original for every input. By the way we defined the
incoming edges of the new neuron $v_{i,t}$, we are guaranteed that for
every input $x$ passed into both $N$ and $\bar{N}$, the value assigned
to $v_{i,t}$ in $\bar{N}$ is greater than the values assigned to both
$v_{i,j}$ and $v_{i,k}$ in the original network. This works to our
advantage, because $v_{i,j}$ and $v_{i,k}$ were both \incN{} --- so
increasing their values increases the output value.  By our 
definition of the outgoing edges, the values of any \incN{} nodes in layer
$i+1$ increase in $\bar{N}$ compared to $N$, and those of any \decN{}
nodes decrease. By definition, this means
that the network's overall output increases.

The abstraction operation for the $\langle$\negN{},\incN{}$\rangle$
case is identical to the one described above. For the remaining two
cases, i.e. $\langle$\posN{},\decN{}$\rangle$ and
$\langle$\negN{},\decN{}$\rangle$, the $\max$ operator in the
definition is replaced with a $\min$ operator.

The next lemma (proof omitted due to lack of space)
justifies the use of our abstraction step, and can be
applied once per each application of \abstractOp{}:
\begin{lemma}
  \label{lemma:correctnessOfAbsract}
  Let $\bar{N}$ be derived from $N$ by a single application of
  \abstractOp{}. For every $x$, it holds that $\bar{N}(x)\geq N(x)$.
\end{lemma}

\subsection{Refinement}
The aforementioned \abstractOp{} operator reduces network size by
merging neurons, but at the cost of accuracy: whereas for some input
$x_0$ the original network returns $N(x_0)=3$, the over-approximation
network $\bar{N}$ created by \abstractOp{} might return
$\bar{N}(x_0)=5$. If our goal is prove that it is never the case that
$N(x)>10$, this over-approximation may be adequate: we can prove that
always $\bar{N}(x)\leq 10$, and this will be enough. However, if our
goal is to prove that it is never the case that $N(x)>4$, the
over-approximation is inadequate: it is possible that the property
holds for $N$, but because $\bar{N}(x_0)=5>4$, our verification
procedure will return $x_0$ as a \emph{spurious counterexample} (a
counterexample for $\bar{N}$ that is not a counterexample for $N$). In
order to handle this situation, we define a \emph{refinement
  operator}, \refineOp{}, that is the inverse of \abstractOp{}: it
transforms $\bar{N}$ into yet another over-approximation, $\bar{N}'$,
with the property that for every $x$,
$N(x)\leq \bar{N}'(x)\leq\bar{N}(x)$. If $\bar{N}'(x_0)=3.5$, it might
be a suitable over-approximation for showing that never $N(x)>4$. In
this section we define the \refineOp{} operator, and in
Section~\ref{sec:cegar} we explain how to use \abstractOp{} and
\refineOp{} as part of a CEGAR-based verification scheme.

Recall that \abstractOp{} merges together a couple of
neurons that share the same attributes. After a series of applications
of \abstractOp{}, each hidden layer $i$ of the resulting network can
be regarded as a partitioning of hidden layer $i$ of the original
network, where each partition contains original, \emph{concrete} neurons that share the same
attributes. In the abstract network, each partition is represented by
a single, \emph{abstract} neuron. The weights on the incoming and outgoing edges
of this abstract neuron are determined according to the definition of the \abstractOp{}
operator. For example, in the case of an abstract neuron $\bar{v}$ that represents a set of concrete neurons $\{v_1,\ldots,v_n\}$ all with attributes
$\langle$\posN{},\incN{}$\rangle$, the weight of each incoming edge to $\bar{v}$ is given by\[
\bar{w}(u,v) = \max(w(u,v_1),\ldots,w(u,v_n))
\]
where $u$ represents a neuron that has not been abstracted yet, and $w$ is the weight function of the original network. The key point here is that the order of \abstractOp{} operations that merged $v_1,\ldots,v_n$ does not matter --- but rather, only the fact that they are now grouped together determines the abstract network's weights.
The following corollary, which is a direct result of Lemma~\ref{lemma:correctnessOfAbsract}, establishes this
connection between sequences of \abstractOp{} applications and
partitions:
\begin{corollary}
  \label{lemma:abstractionsAndPartitions}
  Let $N$ be a DNN where each hidden neuron is labeled as
  \posN{}/\negN{} and \incN{}/\decN{}, and let $\mathcal{P}$ be a
  partitioning of the hidden neurons of $N$, that only groups together
  hidden neurons from the same layer that share the same labels. Then
  $N$ and $\mathcal{P}$ give rise to an abstract neural network
  $\bar{N}$, which is obtained by performing a series of \abstractOp{}
  operations that group together neurons according to the partitions
  of $\mathcal{P}$.  This $\bar{N}$ is an over-approximation of $N$.
\end{corollary}

We now define a \refineOp{} operation that is, in a sense, the inverse
of \abstractOp{}. \refineOp{} takes as input a DNN
$\bar{N}$ that was generated from $N$ via a sequence of \abstractOp{}
operations, and splits a neuron from $\bar{N}$ in two. Formally, 
 the operator receives the original network $N$, the
partitioning $\mathcal{P}$, and a finer partition $\mathcal{P}'$ that
is obtained from $\mathcal{P}$ by splitting a single class in two. The
operator then returns a new abstract network, $\bar{N}'$, that is the
abstraction of $N$ according to $\mathcal{P}'$.

Due to Corollary~\ref{lemma:abstractionsAndPartitions}, and because
$\bar{N}$ returned by \refineOp{} corresponds to a partition
$\mathcal{P}'$ of the hidden neurons of $N$, it is straightforward to
show that $\bar{N}$ is indeed an over-approximation of $N$. The other
useful property that we require is the following:

\begin{lemma}
  \label{lemma:correctnessOfRefine}
  Let $\bar{N}$ be an abstraction of $N$, and let $\bar{N}'$ be a
  network obtained from $\bar{N}$ by applying a single \refineOp{}
  step. Then for every input $x$ it holds that
  $\bar{N}(x)\geq\bar{N}'(x)\geq N(x)$.
\end{lemma}

The second part of the inequality, $\bar{N}'(x)\geq N(x)$ holds
because $\bar{N}'$ is an over-approximation of $N$
(Corollary~\ref{lemma:abstractionsAndPartitions}). The first part of
the inequality, $\bar{N}(x)\geq\bar{N}'(x)$, follows from the fact
that $\bar{N}(x)$ can be obtained from $\bar{N}'(x)$ by a single
application of \abstractOp{}.

In practice, in order to support the refinement of an abstract DNN, we
maintain the current partitioning, i.e. the mapping from concrete
neurons to the abstract neurons that represent them. Then, when an
abstract neuron is selected for refinement (according to some
heuristic, such as the one we propose in Section~\ref{sec:cegar}), we
adjust the mapping and use it to compute the weights of the edges that
touch the affected neuron.

\section{A CEGAR-Based Approach}
\label{sec:cegar}

In Section~\ref{sec:abstractionRefinement} we defined the
\abstractOp{} operator that reduces network size at the cost of reducing network
accuracy, and its inverse \refineOp{} operator that increases network
size and restores accuracy. Together with a black-box verification
procedure \emph{Verify} that can dispatch queries of the form $\varphi
= \langle N, P, Q\rangle$, these components now allow us to design an
abstraction-refinement algorithm for DNN verification, given as Alg.~\ref{alg:abVerification} (we assume that all
hidden neurons in the input network have already been marked \posN{}/\negN{}
and \incN{}/\decN{}).

\begin{algorithm}
  \caption{Abstraction-based DNN Verification($N,P,Q$)}
  \begin{algorithmic}[1]
    \label{alg:cegar}
    \STATE Use \abstractOp{} to generate an initial over-approximation $\bar{N}$ of $N$ \label{step:abstraction}
    \IF {\emph{Verify}($\bar{N},P,Q$) is \unsat{}} \label{step:verify}
      \STATE return \unsat{}
    \ELSE
      \STATE Extract counterexample $c$
      \IF {$c$ is a counterexample for $N$}
        \STATE return \sat{}
      \ELSE
        \STATE Use \refineOp{} to refine $\bar{N}$ into $\bar{N}'$ \label{step:refinement}
        \STATE $\bar{N}\leftarrow \bar{N}'$
        \STATE Goto step~\ref{step:verify}
      \ENDIF
    \ENDIF
  \end{algorithmic}
  \label{alg:abVerification}
\end{algorithm}

Because $\bar{N}$ is obtained via
applications of \abstractOp{} and \refineOp{}, 
the soundness of the underlying \emph{Verify} procedure, together
with Lemmas~\ref{lemma:correctnessOfAbsract} and
\ref{lemma:correctnessOfRefine}, guarantees the soundness of Alg.~\ref{alg:abVerification}. Further, the algorithm always terminates: this is the case
because all the \abstractOp{} steps are performed first, followed by a
sequence of \refineOp{} steps. Because no additional \abstractOp{}
operations are performed beyond Step~\ref{step:abstraction}, after
finitely many \refineOp{} steps $\bar{N}$ will become identical to
$N$, at which point no spurious counterexample will be found, and the
algorithm will terminate with either \sat{} or \unsat{}. Of course,
termination is only guaranteed when the underlying \emph{Verify}
procedure is guaranteed to terminate.

There are two steps in the algorithm that we intentionally left
ambiguous: Step~\ref{step:abstraction}, where the initial
over-approximation is computed, and Step~\ref{step:refinement}, where
the current abstraction is refined due to the discovery of a spurious
counterexample. The motivation was to make Alg.~\ref{alg:cegar}
general, and allow it to be customized by plugging in different heuristics for
performing Steps~\ref{step:abstraction} and~\ref{step:refinement},
which may depend on the problem at hand. Below we propose a few such heuristics.

\subsection{Generating an Initial Abstraction}
The most na\"{i}ve way to generate the initial abstraction is to
apply the \abstractOp{} operator to saturation. As previously
discussed, \abstractOp{} can merge together any pair of hidden neurons
from a given layer that share the same attributes. Since there are four
possible attribute combinations, this will result in each hidden layer
of the network having four neurons or fewer. This method, which we
refer to as \emph{abstraction to saturation}, produces
the smallest abstract networks possible. The downside is that, in some
case, these networks might be too coarse, and might require multiple
rounds of refinement before  a \sat{} or \unsat{} answer can be reached.

A different heuristic for producing abstractions that may require
fewer refinement steps is as follows. First, we select a finite set of
input points, $X=\{x_1,\ldots,x_n\}$, all of which satisfy the input
property $P$. These points can be generated randomly, or 
according to some coverage criterion of the input space.  
The points of $X$ are then used as indicators in estimating when the
abstraction has become too coarse: after every abstraction step, we
check whether the property still holds for $x_1,\ldots,x_n$, and stop
abstracting if this is not the case.  The exact technique, which we
refer to as \emph{indicator-guided abstraction}, appears in
Alg.~\ref{alg:createInitialAbstraction}, which is used to perform
Step~\ref{step:abstraction} of Alg.~\ref{alg:cegar}.

\begin{algorithm}
  \caption{Indicator-Guided Abstraction($N,P,Q,X$)}
  \begin{algorithmic}[1]
    \label{alg:createInitialAbstraction}
    \STATE $\bar{N}\leftarrow N$
    \WHILE {$\forall x\in X.\ \bar{N}(x)$ satisfies $Q$ and there are
      still neurons that can be merged}
    \STATE $\Delta\leftarrow\infty$, bestPair $\leftarrow\bot$
    \FOR {every pair of hidden neurons $v_{i,j}, v_{i,k}$ with
      identical attributes} \label{line:allPairs}
    \STATE m $\leftarrow 0$
    \FOR {every node $v_{i-1,p}$}
    \STATE a $\leftarrow \bar{w}(v_{i-1,p},v_{i,j})$, b $\leftarrow \bar{w}(v_{i-1,p},v_{i,k})$
    \IF {$|a-b|>$ m}
    \STATE m $\leftarrow |a-b|$
    \ENDIF
    \ENDFOR
    \IF {m $<\Delta$}
    \STATE $\Delta\leftarrow$ m, bestPair $\leftarrow \langle v_{i,j}, v_{i,k}\rangle$
    \ENDIF
    \ENDFOR
    \STATE Use \abstractOp{} to merge the nodes of bestPair, store the result in $\bar{N}$
    \ENDWHILE
    \RETURN $\bar{N}$
  \end{algorithmic}
\end{algorithm}

Another point that is addressed by
Alg.~\ref{alg:createInitialAbstraction}, besides how many rounds of
abstraction should be performed, is which pair of neurons should be
merged in every application of \abstractOp{}. This, too, is determined
heuristically. Since any pair of neurons that we pick will result in
the same reduction in network size, our strategy is to prefer neurons
that will result in a more accurate approximation. Inaccuracies are
caused by the $\max{}$ and $\min{}$ operators within the \abstractOp{}
operator: e.g., in the case of $\max{}$, every pair of incoming edges
with weights $a,b$ are replaced by a single edge with weight
$\max{}(a,b)$. Our strategy here is to merge the pair of neurons for
which the \emph{maximal} value of $|a-b|$ (over all incoming edges
with weights $a$ and $b$) is \emph{minimal}. Intuitively, this leads
to $\max{}(a,b)$ being close to both $a$ and $b$ --- which, in turn,
leads to an over-approximation network that is smaller than the
original, but is close to it weight-wise. We point out that although
repeatedly exploring all pairs (line~\ref{line:allPairs}) may appear
costly, in our experiments the time cost of this step was negligible
compared to that of the verification queries that followed. Still, if
this step happens to become a bottleneck, it is possible to adjust the
algorithm to heuristically sample just some of the pairs, and pick the
best pair among those considered --- without harming the algorithm's
soundness.

As a small example, consider the network depicted on the left hand
side of Fig.~\ref{fig:abstraction}. This network has three pairs of
neurons that can be merged using \abstractOp{} (any subset of
$\{v_1,v_2,v_3\}$). Consider the pair $v_1,v_2$: the maximal value of
$|a-b|$ for these neurons is $\max{}(|1-4)|,|(-2)-(-1)|)=3$. For pair
$v_1,v_3$, the maximal value is 1; and for pair $v_2,v_3$ the maximal
value is 2. According to the strategy described in
Alg.~\ref{alg:createInitialAbstraction}, we would first choose to
apply \abstractOp{} on the pair with the minimal maximal value,
i.e. on the pair $v_1, v_3$.

\subsection{Performing the Refinement Step}
\label{sec:refinementSteps}
A refinement step is performed when a spurious counterexample $x$ has
been found, indicating that the abstract network is too coarse. In
other words, our abstraction steps, and specifically the $\max{}$ and
$\min{}$ operators that were used to select edge weights for the
abstract neurons, have resulted in the abstract network's output being
too great for input $x$, and we now need to reduce it. Thus, our
refinement strategies are aimed at applying \refineOp{} in a way that
will result in a significant reduction to the abstract network's
output. We note that there may be multiple options for applying
\refineOp{}, on different nodes, such that any of them would remove
the spurious counterexample $x$ from the abstract network. In
addition, it is not guaranteed that it is possible to remove $x$ with
a single application of \refineOp{}, and multiple consecutive
applications may be required.

One heuristic approach for refinement follows the well-studied notion
of counterexample-guided abstraction
refinement~\cite{ClGrJhLuVe10CEGAR}. Specifically, we leverage the
spurious counterexample $x$ in order to identify a concrete neuron
$v$, which is currently mapped into an abstract neuron $\bar{v}$, such
that splitting $v$ away from $\bar{v}$ might rule out counterexample
$x$. To do this, we evaluate the original network on $x$ and compute
the value of $v$ (we denote this value by $v(x)$), and then do the
same for $\bar{v}$ in the abstract network (value denoted
$\bar{v}(x)$). Intuitively, a neuron pair $\langle v, \bar{v}\rangle$
for which the difference $|v(x) - \bar{v}(x)|$ is significant makes a
good candidate for a refinement operation that will split $v$ away
from $\bar{v}$.

In addition to considering $v(x)$ and $\bar{v}(x)$, we propose to also
consider the weights of the incoming edges of $v$ and $\bar{v}$. When
these weights differ significantly, this could indicate that $\bar{v}$
is too coarse an approximation for $v$, and should be refined. We
argue that by combining these two criteria --- edge weight difference
between $v$ and $\bar{v}$, which is a property of the current abstraction,
together with the difference between $v(x)$ and $\bar{v}(x)$, which is a
property of the specific input $x$, we can identify abstract neurons
that have contributed significantly to $x$ being a spurious
counterexample.

The refinement heuristic is formally defined in
Alg.~\ref{alg:weightBasedRefinement}. The algorithm traverses the
original neurons, looks for the edge weight times assignment value
that has changed the most as a result of the current abstraction, and
then performs refinement on the neuron at the end of that edge. As was
the case with Alg.~\ref{alg:createInitialAbstraction}, if considering
all possible nodes turns out to be too costly, it is possible to
adjust the algorithm to explore only some of the nodes, and pick the
best one among those considered --- without jeopardizing the
algorithm's soundness.

\begin{algorithm}
  \caption{Counterexample-Guided Refinement$(N,\bar{N},x)$}
  \begin{algorithmic}[1]
    \label{alg:weightBasedRefinement}
    \STATE bestNeuron $\leftarrow\bot$, $m \leftarrow 0$
    \FOR {each concrete neuron $v_{i,j}$ of $N$ mapped into abstract neuron $\bar{v}_{i,j'}$ of $\bar{N}$}
    \FOR {each concrete neuron $v_{i-1,k}$ of $N$ mapped into abstract neuron $\bar{v}_{i-1,k'}$ of $\bar{N}$}
    \IF {$|w(v_{i-1,k},v_{i,j}) -
      \bar{w}(\bar{v}_{i-1,k'},\bar{v}_{i,j'})|\cdot |v_{i,j}(x) - \bar{v}_{i,j'}(x)| > m$} \label{line:cegarLine1}
    \STATE $m\leftarrow |w(v_{i-1,k},v_{i,j}) -
    \bar{w}(\bar{v}_{i-1,k'}, \bar{v}_{i,j'})| \cdot |v_{i,j}(x) - \bar{v}_{i,j'}(x)| $ \label{line:cegarLine2}
    \STATE bestNeuron $\leftarrow v_{i,j}$
    \ENDIF
    \ENDFOR
    \ENDFOR
    \STATE Use \refineOp{} to split bestNeuron from its abstract neuron
  \end{algorithmic}
\end{algorithm}

As an example, let us use Alg.~\ref{alg:weightBasedRefinement} to
choose a refinement step for the right hand side network of
Fig.~\ref{fig:abstraction}, for a spurious counterexample $\langle
x_1,x_2\rangle = \langle 1, 0\rangle$. For this input, the original
neurons' evaluation is $v_1=1, v_2=4$ and $v_3=2$, whereas the
abstract neuron that represents them evaluates to $4$.
Suppose $v_1$ is considered first. In the
abstract network, $\bar{w}(x_1,\bar{v_1})=4$ and
$\bar{w}(x_2,\bar{v_1})=-1$; whereas in the original network,
$w(x_1,v_1)=1$ and $w(x_2,v_1)=-2$.
Thus, the largest value $m$
computed for $v_1$ is $|w(x_1,v_1) - \bar{w}(x_1,\bar{v_1})|\cdot|4 -1| =
3\cdot 3 = 9$. This value of $m$ is larger than the one computed for $v_2$ (0)
and for $v_3$ (4), and so $v_1$ is selected for the refinement
step. After this step is performed, $v_2$ and $v_3$ are still mapped
to a single abstract neuron, whereas $v_1$ is mapped to a separate
neuron in the abstract network.

\section{Implementation and Evaluation}
\label{sec:evaluation}
Our implementation of the abstraction-refinement
framework includes modules that read a DNN in the
NNet format~\cite{JuLoBrOwKo16} and a property to be verified, create
an initial abstract DNN as described in Section~\ref{sec:cegar},
invoke a black-box verification engine, and perform refinement as
described in Section~\ref{sec:cegar}. The process terminates when the
underlying engine returns either \unsat{}, or an assignment that is a
true counterexample for the original network.  For experimentation
purposes, we integrated our framework with the Marabou DNN
verification engine~\cite{KaHuIbJuLaLiShThWuZeDiKoBa19Marabou}.  Our
implementation and benchmarks are publicly available online~\cite{cegarabouCode}.

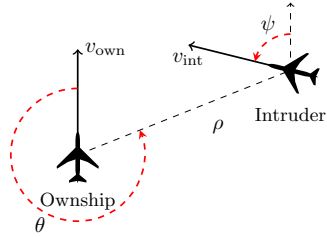
\begin{wrapfigure}[13]{r}{5.8cm}
  \vspace{-1.2cm}
  \begin{center}
	\scalebox{0.8}{
		\def\layersep{1.6cm}
\def\layerseP{1.12cm}

\begin{tikzpicture}[scale=0.7]

\node[aircraft top,fill=black,minimum width=1cm,rotate=90,scale = 0.85] (Own) at (0,0) {} node [below,yshift=-0.5cm,font=\footnotesize] {Ownship};
\coordinate[label=right:$v_\text{own}$] (S0) at (0,2.5);

\node[aircraft top,fill=black,minimum width=1cm,rotate=166, scale = 0.85] (Int) at (5,2) {};
\node at (5,2) [below,yshift=-0.5cm,font=\footnotesize] {Intruder};
\coordinate[label=above:$$] (IntN) at (5,3.6);
\coordinate[label=below:$v_\text{int}$] (S1) at (2.6,2.6);

\coordinate[label=below:$\rho$] (R) at (3.3,1.0);

\draw [thick, ->] (Own) -- (S0);
\draw [dashed,->] (Int) -- (IntN);
\draw [thick, ->] (Int) -- (S1);
            
\draw[dashed] (Own) -- (Int);

\pic [draw,-stealth,red,thick,dashed,angle radius=0.6cm,"$\psi$"{anchor=west,text = black, above}, angle eccentricity=1] {angle = IntN--Int--S1};
\pic [draw,-stealth,red,thick,dashed,angle radius=1.1cm,"$\theta$"{anchor=west,text = black, below}, angle eccentricity=1] {angle = S0--Own--Int};

\end{tikzpicture}
	}
	\caption{(From~\cite{KaBaDiJuKo17Reluplex}) An illustration of
          the sensor readings
          passed as input to the ACAS Xu DNNs.}
	\label{fig:acasXuGeometry}
  \end{center}
\end{wrapfigure}
      
Our experiments included verifying several
properties of the 45 ACAS Xu DNNs for airborne collision
avoidance~\cite{JuLoBrOwKo16,KaBaDiJuKo17Reluplex}. ACAS Xu is a
system designed to produce horizontal turning advisories for an
unmanned aircraft (the \emph{ownship}), with the purpose of preventing
a collision with another nearby aircraft (the \emph{intruder}). The
ACAS Xu system receive as input sensor readings, indicating the
location of the intruder relative to the ownship, the speeds of the
two aircraft, and their directions (see
Fig.~\ref{fig:acasXuGeometry}). Based on these readings, it
selects one of 45 DNNs, to which the readings are then passed as input.
The selected DNN then assigns scores to five output neurons, each
representing a possible turning advisory: strong left, weak left,
strong right, weak right, or clear-of-conflict (the latter indicating
that it is safe to continue along the current trajectory). The neuron
with the \emph{lowest} score represents the selected advisory. We
verified several properties of these DNNs based on the list of
properties that appeared in~\cite{KaBaDiJuKo17Reluplex} ---
specifically focusing on properties that ensure that the DNNs always advise
clear-of-conflict for distant intruders, and that they are robust to
(i.e., do not change their advisories in the presence of)
small input perturbations.

Each of the ACAS Xu DNNs has 300 hidden nodes spread across 6 hidden
layers, leading to 1200 neurons when the transformation from
Section~\ref{sec:abstraction} is applied. In our experiments we set
out to check whether the abstraction-based approach could indeed prove
properties of the ACAS Xu networks on abstract networks that had
significantly fewer neurons than the original ones. In addition, we
wished to compare the proposed approaches for generating initial
abstractions (the abstraction to saturation approach versus the
indicator-guided abstraction described in
Alg.~\ref{alg:createInitialAbstraction}), in order to identify an
optimal configuration for our tool. Finally, once the optimal
configuration has been identified, we used it to compare our tool's
performance to that of vanilla Marabou. The results are described
next.

\begin{figure}[h]
  \begin{center}
    \scalebox{0.8}{
      \hspace{-0.5cm}
      \includegraphics[width=1.32\textwidth]{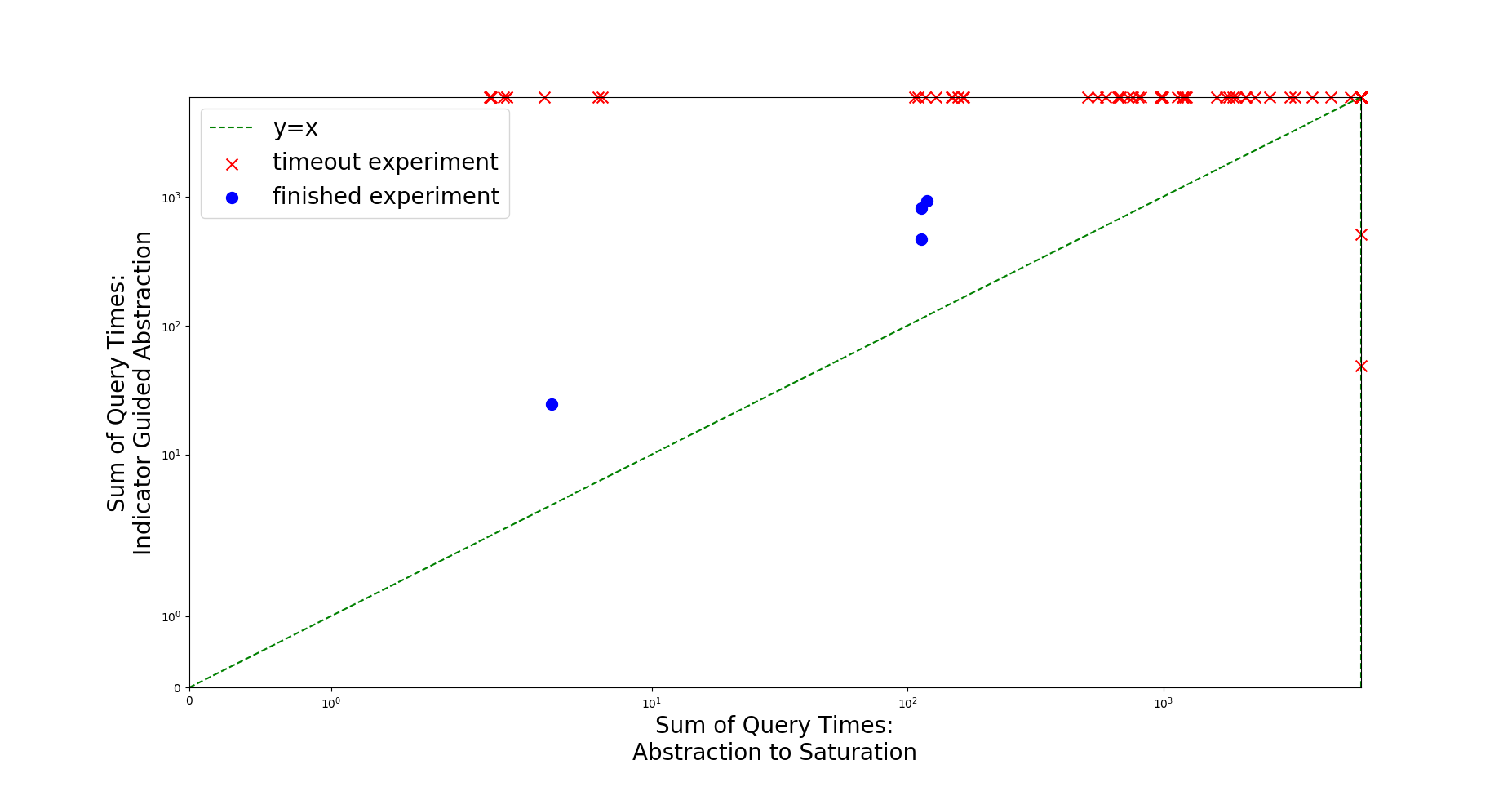}
      }
  \end{center}
  \caption{Generating initial abstractions using abstraction to
    saturation and indicator-guided abstraction.}
  \label{fig:heursticVsComplete}
\end{figure}

Fig.~\ref{fig:heursticVsComplete} depicts a comparison of the two
approaches for generating initial abstractions: the abstraction to
saturation scheme (x axis), and the indicator-guided abstraction
scheme described in Alg.~\ref{alg:createInitialAbstraction} (y
axis). Each experiment included running our tool twice on the same
benchmark (network and property), with an identical configuration
except for the initial abstraction being used. The plot depicts the
total time (log-scale, in seconds, with a 20-hour timeout) spent by
Marabou solving verification queries as part of the
abstraction-refinement procedure. It shows that, in contrast to our
intuition, abstraction to saturation almost always outperforms the
indicator-guided approach. This is perhaps due to the fact that,
although it might entail additional rounds of refinement, the
abstraction to saturation approach tends to produce coarse verification
queries that are easily solved by Marabou,
resulting in an overall improved performance.  We thus conclude that,
at least in the ACAS Xu case, the abstraction to saturation approach
is superior to that of indicator-guided abstraction.

 This experiment also confirms that properties can indeed be proved on
abstract networks that are significantly smaller than the original ---
i.e., despite the initial 4x increase in network size due to the
preprocessing phase, the final abstract network on which our
abstraction-enhanced approach could solve the query was usually
substantially smaller than the original network. Specifically, among
the abstraction to saturation experiments that terminated, the final
network on which the property was shown to be \sat{} or \unsat{} had
an average size of 268.8 nodes, compared to the original 310 --- a
13\% reduction. Because DNN verification becomes exponentially more
difficult as the network size increases, this reduction is highly
beneficial.

\begin{figure}[htp]
  \begin{center}
    \scalebox{0.8}{
            \hspace{-0.5cm}
        \includegraphics[width=1.32\textwidth]{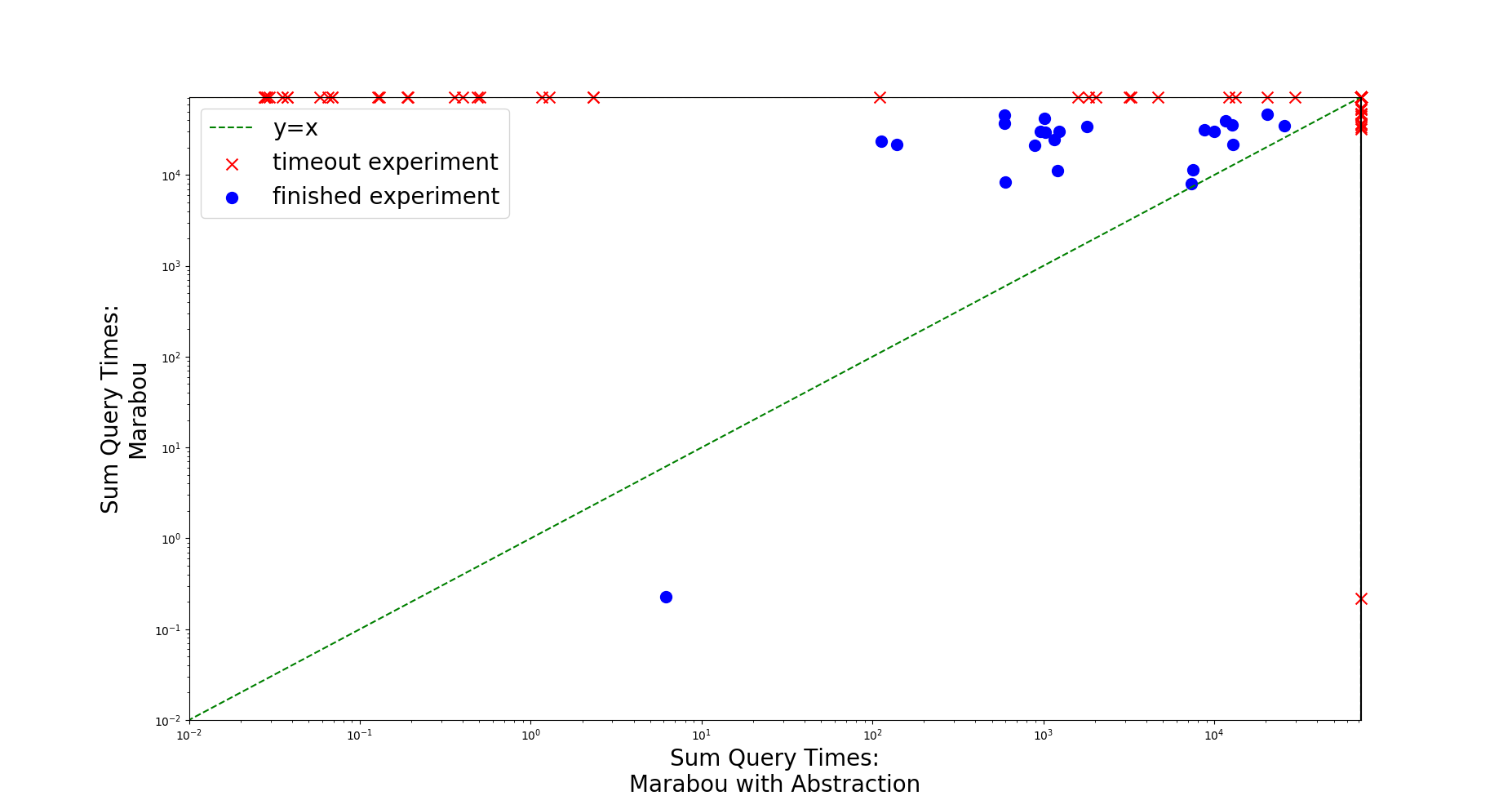}
      }
  \end{center}
  \caption{Comparing the run time (in seconds, logscale) of vanilla Marabou and
    the abstraction-enhanced version on the ACAS Xu benchmarks.}
  \label{fig:cegarVsVanila}
\end{figure}

Next, we compared our abstraction-enhanced Marabou (in abstraction to
saturation mode) to the vanilla version.  The plot in
Fig.~\ref{fig:cegarVsVanila} compares the total query solving time of
vanilla Marabou (y axis) to that of our approach (x axis). We ran the
tools on $90$ ACAS Xu benchmarks (2 properties, checked on each of the
45 networks), with a 20-hour timeout.  We observe that the
abstraction-enhanced version significantly outperforms vanilla Marabou
on average --- often solving queries orders-of-magnitude more quickly,
and timing out on fewer benchmarks. Specifically, the
abstraction-enhanced version solved 58 instances, versus 35 solved by
Marabou. Further, over the instances solved by both tools, the
abstraction-enhanced version had a total query median runtime of 1045
seconds, versus 63671 seconds for Marabou. Interestingly, the average
size of the abstract networks for which our tool was able to solve the
query was 385 nodes --- which is an increase compared to the original
310 nodes. However, the improved runtimes demonstrate that although
these networks were slightly larger, they were still much easier to
verify, presumably because many of the network's original neurons
remained abstracted away.

Finally, we used our abstraction-enhanced Marabou to verify
\emph{adversarial robustness}
properties~\cite{SzZaSuBrErGoFe13}. Intuitively, an adversarial
robustness property states that slight input perturbations cannot
cause sudden spikes in the network's output. This is desirable because
such sudden spikes can lead to misclassification of inputs. Unlike the
ACAS Xu domain-specific properties~\cite{KaBaDiJuKo17Reluplex}, whose
formulation required input from human experts, adversarial robustness
is a \emph{universal property}, desirable for every DNN. Consequently
it is easier to formulate, and has received much attention
(e.g.,~\cite{BaIoLaVyNoCr16,GeMiDrTsChVe18,KaBaDiJuKo17Reluplex,TjXiTe19}).

In order to formulate adversarial robustness properties for the ACAS
Xu networks, we randomly sampled the ACAS Xu DNNs to identify input
points where the selected output advisory, indicated by an output
neuron $y_i$, received a much lower score than the second-best
advisory, $y_j$ (recall that the advisory with the lowest score is
selected). For such an input point $x_0$, we then posed the
verification query: does there exist a point $x$ that is close to
$x_0$, but for which $y_j$ receives a lower score than $y_i$? Or, more
formally:
$
\left(  \lVert x - x_0 \rVert_{L_\infty}\leq \delta\right) \wedge (y_j \leq y_i).
$
 If this query is \sat{} then there exists an input $x$
whose distance to $x_0$ is at most $\delta$, but for which the network
assigns a better (lower) score to advisory $y_j$ than to
$y_i$. However, if this query is \unsat{}, 
no such point $x$ exists. Because we select point $x_0$ such that
$y_i$ is initially much smaller than $y_j$, we expect the query to be
\unsat{} for small values of $\delta$.

\begin{figure}[htp]
  \begin{center}
    \scalebox{0.8}{
            \hspace{-0.5cm}
        \includegraphics[width=1.32\textwidth]{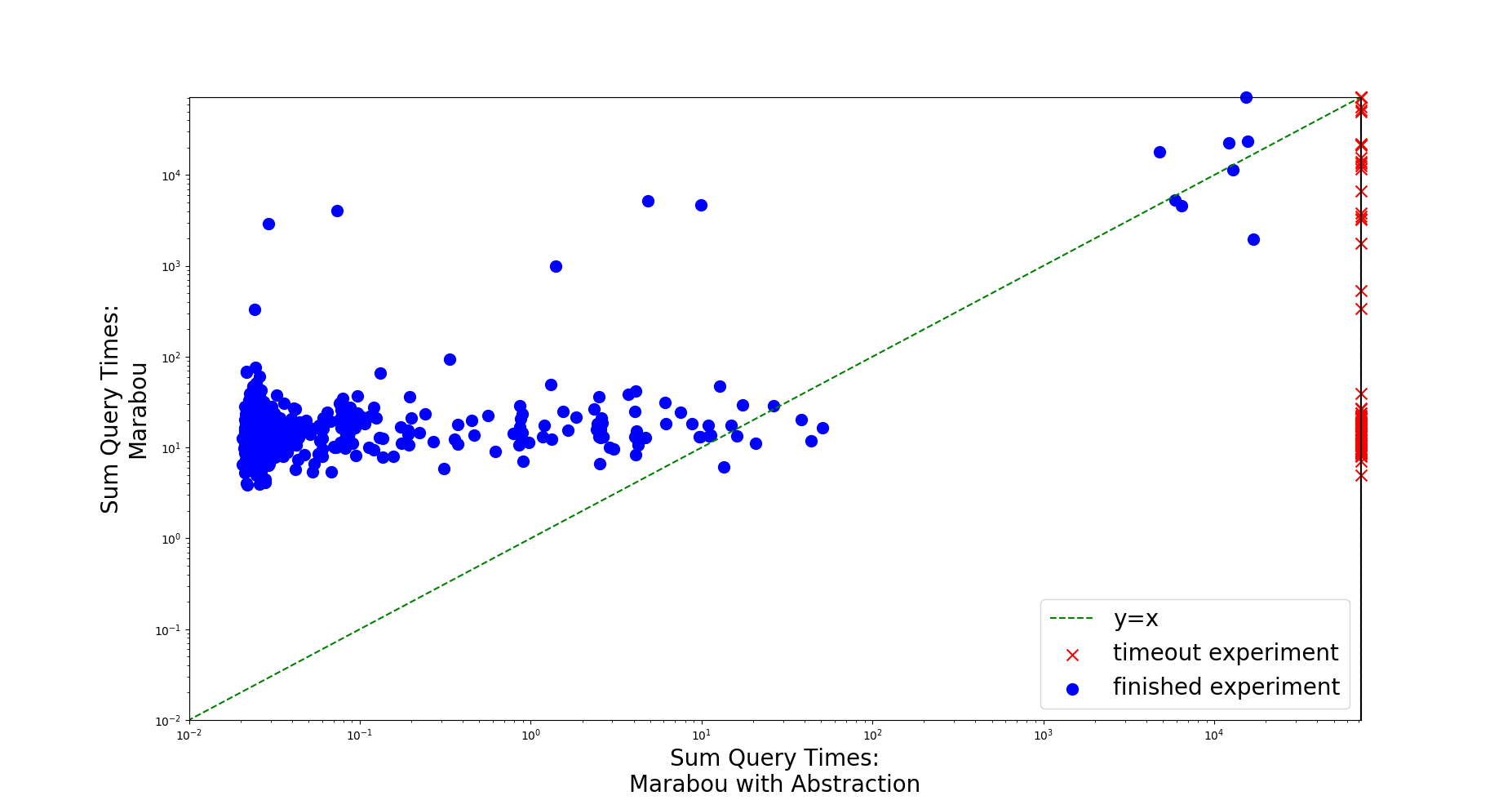}
      }
  \end{center}
  \caption{Comparing the run time (seconds, logscale) of vanilla Marabou and
    the abstraction-enhanced version on the ACAS Xu adversarial
    robustness properties.}
  \label{fig:adversarialRobustness}
\end{figure}

For each of the 45 ACAS Xu networks, we created robustness queries for
$20$ distinct input points --- producing a total of 900 verification
queries (we arbitrarily set $\delta=0.1$). For each of these queries
we compared the runtime of vanilla Marabou to that of our
abstraction-enhanced version (with a 20-hour timeout).  The results
are depicted in Fig.~\ref{fig:adversarialRobustness}. Vanilla Marabou
was able to solve more instances --- 893 out of 900, versus 805 that
the abstraction-enhanced version was able to solve. However, on the
vast majority of the remaining experiments, the abstraction-enhanced
version was significantly faster, with a total query median runtime of
only 0.026 seconds versus 15.07 seconds in the vanilla version (over
the 805 benchmarks solved by both tools). This impressive 99\%
improvement in performance highlights the usefulness of our approach
also in the context of adversarial robustness. In addition, over the
solved benchmarks, the average size of the abstract networks for which
our tool was able to solve the query was 104.4 nodes, versus 310 nodes
in each of the original networks --- a 66\% reduction in size. This
reinforces our statement that, in many cases, DNNs contain a great
deal of unneeded neurons, which can safely be removed by the
abstraction process for the purpose of verification.

\section{Related Work}
\label{sec:relatedWork}
In recent years, multiple schemes have been proposed for the
verification of neural networks.  These include SMT-based approaches,
such as Marabou~\cite{KaHuIbJuLaLiShThWuZeDiKoBa19Marabou,KaBaKaSc19},
Reluplex~\cite{KaBaDiJuKo17Reluplex}, DLV~\cite{HuKwWaWu17} and
others; approaches based on formulating the problem as a mixed integer
linear programming instance
(e.g.,~\cite{BuTuToKoKu17,DuJhSaTi18,Ehlers2017,TjXiTe19}); approaches
that use sophisticated symbolic interval
propagation~\cite{WaPeWhYaJa18}, or abstract
interpretation~\cite{GeMiDrTsChVe18}; and others
(e.g.,~\cite{AnPaDiCh19,JaBaKa20,KuKaGoJuBaKo18,LoMa17,NaKaRySaWa17,WuOzZeIrJuGoFoKaPaBa20,XiTrJo18}). These
approaches have been applied in a variety of tasks, such as measuring
adversarial robustness~\cite{BaIoLaVyNoCr16,HuKwWaWu17}, neural
network simplification~\cite{GoFeMaBaKa20}, neural network
modification~\cite{GoAdKeKa20}, and many others
(e.g.,~\cite{KaBaKaSc19,SuKhSh19}).  Our approach can be integrated
with any sound and complete solver as its engine, and then applied
towards any of the aforementioned tasks.  Incomplete solvers could
also be used and might afford better performance, but this could result in
our approach also becoming incomplete. 

Some existing DNN verification techniques incorporate abstraction
elements. In~\cite{PuTa10}, the authors use abstraction to
over-approximate the Sigmoid activation function with a collection of
rectangles. If the abstract verification query they
produce is \unsat{}, then so is the original. When a spurious
counterexample is found, an arbitrary refinement step is
performed. The authors report limited scalability, tackling only
networks with a few dozen neurons. Abstraction techniques also appear
in the AI2 approach~\cite{GeMiDrTsChVe18}, but there it is the input
property and reachable regions that are over-approximated, as opposed
to the DNN itself. Combining this kind of input-focused abstraction with our
network-focused abstraction is an interesting avenue for future work.

\section{Conclusion}
\label{sec:conclusion}

With deep neural networks becoming widespread and with their forthcoming
integration into safety-critical systems, there is an urgent need for
scalable techniques to verify and reason about them. However, the size
of these networks poses a serious challenge. Abstraction-based
techniques can mitigate this difficulty, by replacing networks with
smaller versions thereof to be verified, without compromising the
soundness of the verification procedure. The abstraction-based
approach we have proposed here can provide a significant reduction in
network size, thus boosting the performance of existing verification
technology.

In the future, we plan to continue this work along several
axes. First, we intend to investigate refinement heuristics that can
split an abstract neuron into two arbitrary sized neurons.  In addition, we
will investigate abstraction schemes for networks that use additional
activation functions, beyond ReLUs. Finally, we plan to make our
abstraction scheme parallelizable, allowing users to use multiple
worker nodes to explore different combinations of abstraction and refinement
steps, hopefully leading to faster convergence.

\medskip
\noindent
\textbf{Acknowledgements.}  We thank the anonymous reviewers for their
insightful comments. This project was partially supported by grants
from the Binational Science Foundation (2017662) and the Israel
Science Foundation (683/18).

\bibliographystyle{abbrv}

\begin{thebibliography}{10}

\bibitem{AnPaDiCh19}
G.~Anderson, S.~Pailoor, I.~Dillig, and S.~Chaudhuri.
\newblock {Optimization and Abstraction: a Synergistic Approach for Analyzing
  Neural Network Robustness}.
\newblock In {\em Proc. 40th ACM SIGPLAN Conf. on Programming Language Design
  and Implementation (PLDI)}, pages 731--744, 2019.

\bibitem{BaIoLaVyNoCr16}
O.~Bastani, Y.~Ioannou, L.~Lampropoulos, D.~Vytiniotis, A.~Nori, and
  A.~Criminisi.
\newblock {Measuring Neural Net Robustness with Constraints}.
\newblock In {\em Proc. 30th Conf. on Neural Information Processing Systems
  (NIPS)}, 2016.

\bibitem{BoDeDwFiFlGoJaMoMuZhZhZhZi16}
M.~Bojarski, D.~Del~Testa, D.~Dworakowski, B.~Firner, B.~Flepp, P.~Goyal,
  L.~Jackel, M.~Monfort, U.~Muller, J.~Zhang, X.~Zhang, J.~Zhao, and K.~Zieba.
\newblock {End to End Learning for Self-Driving Cars}, 2016.
\newblock Technical Report. \url{http://arxiv.org/abs/1604.07316}.

\bibitem{BuTuToKoKu17}
R.~Bunel, I.~Turkaslan, P.~Torr, P.~Kohli, and M.~Kumar.
\newblock {Piecewise Linear Neural Network Verification: A Comparative Study},
  2017.
\newblock Technical Report. \url{https://arxiv.org/abs/1711.00455v1}.

\bibitem{CaKaBaDi17}
N.~Carlini, G.~Katz, C.~Barrett, and D.~Dill.
\newblock {Provably Minimally-Distorted Adversarial Examples}, 2017.
\newblock Technical Report. \url{https://arxiv.org/abs/1709.10207}.

\bibitem{ClGrJhLuVe10CEGAR}
E.~Clarke, O.~Grumberg, S.~Jha, Y.~Lu, and H.~Veith.
\newblock {Counterexample-Guided Abstraction Refinement}.
\newblock In {\em Proc. 12th Int. Conf. on Computer Aided Verification (CAV)},
  pages 154--169, 2010.

\bibitem{DuJhSaTi18}
S.~Dutta, S.~Jha, S.~Sanakaranarayanan, and A.~Tiwari.
\newblock {Output Range Analysis for Deep Neural Networks}.
\newblock In {\em Proc. 10th NASA Formal Methods Symposium (NFM)}, pages
  121--138, 2018.

\bibitem{Ehlers2017}
R.~Ehlers.
\newblock {Formal Verification of Piece-Wise Linear Feed-Forward Neural
  Networks}.
\newblock In {\em Proc. 15th Int. Symp. on Automated Technology for
  Verification and Analysis (ATVA)}, pages 269--286, 2017.

\bibitem{cegarabouCode}
Y.~Y. Elboher, J.~Gottschlich, and G.~Katz.
\newblock {An Abstraction-Based Framework for Neural Network Verification:
  Proof-of-Concept Implementation}.
\newblock
  \url{https://drive.google.com/file/d/1KCh0vOgcOR2pSbGRdbtAQTmoMHAFC2Vs/view},
  2020.

\bibitem{GeMiDrTsChVe18}
T.~Gehr, M.~Mirman, D.~Drachsler-Cohen, E.~Tsankov, S.~Chaudhuri, and
  M.~Vechev.
\newblock {AI2: Safety and Robustness Certification of Neural Networks with
  Abstract Interpretation}.
\newblock In {\em Proc. 39th IEEE Symposium on Security and Privacy (S\&P)},
  2018.

\bibitem{GoFeMaBaKa20}
S.~Gokulanathan, A.~Feldsher, A.~Malca, C.~Barrett, and G.~Katz.
\newblock {Simplifying Neural Networks using Formal Verification}.
\newblock In {\em Proc. 12th NASA Formal Methods Symposium (NFM)}, 2020.

\bibitem{GoAdKeKa20}
B.~Goldberger, Y.~Adi, J.~Keshet, and G.~Katz.
\newblock {Minimal Modifications of Deep Neural Networks using Verification}.
\newblock In {\em Proc. 23rd Int. Conf. on Logic for Programming, Artificial
  Intelligence and Reasoning (LPAR)}, 2020.

\bibitem{FoBeCu16}
I.~Goodfellow, Y.~Bengio, and A.~Courville.
\newblock {\em {Deep Learning}}.
\newblock MIT Press, 2016.

\bibitem{GoKaPaBa18}
D.~Gopinath, G.~Katz, C.~P\v{a}s\v{a}reanu, and C.~Barrett.
\newblock {DeepSafe: A Data-driven Approach for Assessing Robustness of Neural
  Networks}.
\newblock In {\em Proc. 16th. Int. Symp. on on Automated Technology for
  Verification and Analysis (ATVA)}, pages 3--19, 2018.

\bibitem{gottschlich:2018:mapl}
J.~Gottschlich, A.~Solar-Lezama, N.~Tatbul, M.~Carbin, M.~Rinard, R.~Barzilay,
  S.~Amarasinghe, J.~Tenenbaum, and T.~Mattson.
\newblock {The Three Pillars of Machine Programming}.
\newblock In {\em Proc. 2nd ACM SIGPLAN Int. Workshop on Machine Learning and
  Programming Languages (MALP)}, pages 69--80, 2018.

\bibitem{HiDeYuDaMoJaSeVaNgSaKi12}
G.~Hinton, L.~Deng, D.~Yu, G.~Dahl, A.~Mohamed, N.~Jaitly, A.~Senior,
  V.~Vanhoucke, P.~Nguyen, T.~Sainath, and B.~Kingsbury.
\newblock {Deep Neural Networks for Acoustic Modeling in Speech Recognition:
  The Shared Views of Four Research Groups}.
\newblock {\em IEEE Signal Processing Magazine}, 29(6):82--97, 2012.

\bibitem{HuKwWaWu17}
X.~Huang, M.~Kwiatkowska, S.~Wang, and M.~Wu.
\newblock {Safety Verification of Deep Neural Networks}.
\newblock In {\em Proc. 29th Int. Conf. on Computer Aided Verification (CAV)},
  pages 3--29, 2017.

\bibitem{JaBaKa20}
Y.~Jacoby, C.~Barrett, and G.~Katz.
\newblock {Verifying Recurrent Neural Networks using Invariant Inference},
  2020.
\newblock Technical Report. \url{http://arxiv.org/abs/2004.02462}.

\bibitem{JuLoBrOwKo16}
K.~Julian, J.~Lopez, J.~Brush, M.~Owen, and M.~Kochenderfer.
\newblock {Policy Compression for Aircraft Collision Avoidance Systems}.
\newblock In {\em Proc. 35th Digital Avionics Systems Conf. (DASC)}, pages
  1--10, 2016.

\bibitem{KaBaDiJuKo17Reluplex}
G.~Katz, C.~Barrett, D.~Dill, K.~Julian, and M.~Kochenderfer.
\newblock {Reluplex: An Efficient SMT Solver for Verifying Deep Neural
  Networks}.
\newblock In {\em Proc. 29th Int. Conf. on Computer Aided Verification (CAV)},
  pages 97--117, 2017.

\bibitem{KaBaDiJuKo17}
G.~Katz, C.~Barrett, D.~Dill, K.~Julian, and M.~Kochenderfer.
\newblock {Towards Proving the Adversarial Robustness of Deep Neural Networks}.
\newblock In {\em Proc. 1st Workshop on Formal Verification of Autonomous
  Vehicles (FVAV)}, pages 19--26, 2017.

\bibitem{KaHuIbJuLaLiShThWuZeDiKoBa19Marabou}
G.~Katz, D.~Huang, D.~Ibeling, K.~Julian, C.~Lazarus, R.~Lim, P.~Shah,
  S.~Thakoor, H.~Wu, A.~Zelji\'c, D.~Dill, M.~Kochenderfer, and C.~Barrett.
\newblock {The Marabou Framework for Verification and Analysis of Deep Neural
  Networks}.
\newblock In {\em Proc. 31st Int. Conf. on Computer Aided Verification (CAV)},
  2019.

\bibitem{KaBaKaSc19}
Y.~Kazak, C.~Barrett, G.~Katz, and M.~Schapira.
\newblock {Verifying Deep-RL-Driven Systems}.
\newblock In {\em Proc. 1st ACM SIGCOMM Workshop on Network Meets AI \& ML
  (NetAI)}, 2019.

\bibitem{KrSuHi12}
A.~Krizhevsky, I.~Sutskever, and G.~Hinton.
\newblock {Imagenet Classification with Deep Convolutional Neural Networks}.
\newblock {\em Advances in Neural Information Processing Systems}, pages
  1097--1105, 2012.

\bibitem{KuKaGoJuBaKo18}
L.~Kuper, G.~Katz, J.~Gottschlich, K.~Julian, C.~Barrett, and M.~Kochenderfer.
\newblock {Toward Scalable Verification for Safety-Critical Deep Networks},
  2018.
\newblock Technical Report. \url{https://arxiv.org/abs/1801.05950}.

\bibitem{KuGoBe16}
A.~Kurakin, I.~Goodfellow, and S.~Bengio.
\newblock {Adversarial Examples in the Physical World}, 2016.
\newblock Technical Report. \url{http://arxiv.org/abs/1607.02533}.

\bibitem{LoMa17}
A.~Lomuscio and L.~Maganti.
\newblock {An Approach to Reachability Analysis for Feed-Forward ReLU Neural
  Networks}, 2017.
\newblock Technical Report. \url{https://arxiv.org/abs/1706.07351}.

\bibitem{MaNeAl17Pensieve}
H.~Mao, R.~Netravali, and M.~Alizadeh.
\newblock {Neural Adaptive Video Streaming with Pensieve}.
\newblock In {\em Proc. Conf. of the ACM Special Interest Group on Data
  Communication (SIGCOMM)}, pages 197--210, 2017.

\bibitem{NaHi10}
V.~Nair and G.~Hinton.
\newblock {Rectified Linear Units Improve Restricted Boltzmann Machines}.
\newblock In {\em Proc. 27th Int. Conf. on Machine Learning (ICML)}, pages
  807--814, 2010.

\bibitem{NaKaRySaWa17}
N.~Narodytska, S.~Kasiviswanathan, L.~Ryzhyk, M.~Sagiv, and T.~Walsh.
\newblock {Verifying Properties of Binarized Deep Neural Networks}, 2017.
\newblock Technical Report. \url{http://arxiv.org/abs/1709.06662}.

\bibitem{PuTa10}
L.~Pulina and A.~Tacchella.
\newblock {An Abstraction-Refinement Approach to Verification of Artificial
  Neural Networks}.
\newblock In {\em Proc. 22nd Int. Conf. on Computer Aided Verification (CAV)},
  pages 243--257, 2010.

\bibitem{RuHuKw18}
W.~Ruan, X.~Huang, and M.~Kwiatkowska.
\newblock {Reachability Analysis of Deep Neural Networks with Provable
  Guarantees}.
\newblock In {\em Proc. 27th Int. Joing Conf. on Artificial Intelligence
  (IJACI)}, pages 2651--2659, 2018.

\bibitem{SiHuMaGuSiVaScAnPaLaDi16}
D.~Silver, A.~Huang, C.~Maddison, A.~Guez, L.~Sifre, G.~Van Den~Driessche,
  J.~Schrittwieser, I.~Antonoglou, V.~Panneershelvam, M.~Lanctot, and
  S.~Dieleman.
\newblock {Mastering the Game of Go with Deep Neural Networks and Tree Search}.
\newblock {\em Nature}, 529(7587):484--489, 2016.

\bibitem{SuKhSh19}
X.~Sun, K.~H., and Y.~Shoukry.
\newblock {Formal Verification of Neural Network Controlled Autonomous
  Systems}.
\newblock In {\em Proc. 22nd ACM Int. Conf. on Hybrid Systems: Computation and
  Control (HSCC)}, 2019.

\bibitem{SzZaSuBrErGoFe13}
C.~Szegedy, W.~Zaremba, I.~Sutskever, J.~Bruna, D.~Erhan, I.~Goodfellow, and
  R.~Fergus.
\newblock {Intriguing Properties of Neural Networks}, 2013.
\newblock Technical Report. \url{http://arxiv.org/abs/1312.6199}.

\bibitem{TjXiTe19}
V.~Tjeng, K.~Xiao, and R.~Tedrake.
\newblock {Evaluating Robustness of Neural Networks with Mixed Integer
  Programming}.
\newblock In {\em Proc. 7th Int. Conf. on Learning Representations (ICLR)},
  2019.

\bibitem{WaPeWhYaJa18}
S.~Wang, K.~Pei, J.~Whitehouse, J.~Yang, and S.~Jana.
\newblock {Formal Security Analysis of Neural Networks using Symbolic
  Intervals}.
\newblock In {\em Proc. 27th USENIX Security Symposium}, 2018.

\bibitem{WuOzZeIrJuGoFoKaPaBa20}
H.~Wu, A.~Ozdemir, A.~Zelji\'c, A.~Irfan, K.~Julian, D.~Gopinath, S.~Fouladi,
  G.~Katz, C.~P\u{a}s\u{a}reanu, and C.~Barrett.
\newblock {Parallelization Techniques for Verifying Neural Networks}, 2020.
\newblock Technical Report. \url{https://arxiv.org/abs/2004.08440}.

\bibitem{XiTrJo18}
W.~Xiang, H.-D. Tran, and T.~Johnson.
\newblock {Output Reachable Set Estimation and Verification for Multilayer
  Neural Networks}.
\newblock {\em IEEE Transactions on Neural Networks and Learning Systems
  (TNNLS)}, 99:1--7, 2018.

\end{thebibliography}

\end{document}